\pdfoutput=1
\documentclass[aps,prd,amsmath,floats,floatfix, twocolumn,
superscriptaddress,nofootinbib,showpacs,longbibliography]{revtex4-1}

\usepackage[T1]{fontenc}
\usepackage[utf8]{inputenc}
\usepackage{lmodern}

\usepackage{verbatim}
\usepackage{makecell}

\usepackage[dvipsnames, usenames]{xcolor}
\definecolor{linkcolor}{rgb}{0.0,0.3,0.5}
\usepackage[hypertexnames=false, unicode, colorlinks=true, linkcolor=linkcolor,
citecolor=linkcolor, filecolor=linkcolor,urlcolor=linkcolor,
pdfusetitle]{hyperref}
\usepackage[all]{hypcap}
\usepackage{graphicx}
\usepackage{xspace}
\usepackage{braket}
\usepackage{amssymb}
\usepackage[normalem]{ulem} 
\usepackage{bm} 

\usepackage{array}
\usepackage{multirow}
\usepackage{microtype}

\usepackage{lipsum}  

\usepackage{blindtext}
\usepackage{mathrsfs}
\usepackage{orcidlink}
\graphicspath{%
  {figs/}%
}

\DeclareMathAlphabet{\mathpzc}{OT1}{pzc}{m}{it}

\newcommand{\UMassDMath}{\affiliation{Department of Mathematics, University of Massachusetts, Dartmouth, MA 02747, USA}}
\newcommand{\UMassDPhysics}{\affiliation{Department of Physics, University of Massachusetts, Dartmouth, MA 02747, USA}}
\newcommand{\CSCDRUMass}{\affiliation{Center for Scientific Computing and Data Science Research, University of Massachusetts, Dartmouth, MA 02747, USA}}
\newcommand{\URI}{\affiliation{Department of Physics and Center for Computational Research, University of Rhode Island, Kingston, RI 02881.}}
\newcommand{\UTAustin}{\affiliation{Center for Gravitational Physics, University of Texas, Austin, TX 78712.}}

\newcommand{\TITLE}{Toward exponentially-convergent simulations of extreme-mass-ratio inspirals: \\ A time-domain solver for the scalar Teukolsky equation with singular source terms}

\begin{document}

\title{\TITLE}

\author{Manas Vishal\,\orcidlink{0000-0003-3424-3505}} \email{vishalmanas28@gmail.com} \UMassDMath \CSCDRUMass
\author{Scott E. Field\,\orcidlink{0000-0002-6037-3277}} \UMassDMath \CSCDRUMass
\author{Katie Rink\,\orcidlink{0000-0002-1494-3494}} \UMassDPhysics \CSCDRUMass \UTAustin
\author{Sigal Gottlieb\,\orcidlink{0000-0002-6526-3886}} \UMassDMath \CSCDRUMass
\author{Gaurav Khanna} \URI \UMassDPhysics \CSCDRUMass

\date{\today}

\begin{abstract}
Gravitational wave signals from extreme mass ratio inspirals are a key target for upcoming, space-based gravitational wave detectors. These systems are typically modeled as a distributionally-forced Teukolsky equation, where the smaller black hole is treated as a Dirac delta distribution (i.e., a point-particle). Time-domain solvers often use regularization approaches that approximate the Dirac distribution. Unfortunately, such approaches often introduce small length scales (e.g., when the approximation is by a narrow Gaussian) and are a source of systematic error, especially near the smaller black hole. We describe a multi-domain discontinuous Galerkin (DG) method for solving the distributionally-forced $s=0$ Teukolsky equation that describes scalar fields evolving on a Kerr spacetime. To handle the Dirac delta, we expand the solution in spherical harmonics and recast the sourced Teukolsky equation as a first-order, one-dimensional symmetric hyperbolic system. This allows us to derive the method's numerical flux to correctly account for the Dirac delta. As a result, our method achieves global spectral accuracy even at the source's location. To connect the near field to future null infinity, we use the hyperboloidal layer method, allowing us to supply trivial outer boundary conditions and providing direct access to the far-field waveform. We document several numerical experiments where we test our method, including convergence tests against exact solutions, energy luminosities for circular orbits, the scheme's superconvergence properties at future null infinity, and the late-time tail behavior of the scalar field. We also compare two systems that arise from different choices of the first-order reduction variables, finding that certain reasonable choices are numerically problematic in practice. The methods developed here may be beneficial when computing gravitational self-force effects, where the regularization procedure has been developed for the spherical harmonic modes and high accuracy is needed at the location of the Dirac delta.
\end{abstract}

\maketitle


\section{Introduction}
\label{Sec:Introduction}

Black hole perturbation theory is a standard framework for studying a diverse range of gravitational phenomena, such as gravitational waves, quasi-normal modes, late-time Price tails, self-force effects, and linear stability of black hole solutions. The theory of such perturbations started with pioneering investigations by Regge and Wheeler~\cite{Regge:1957td} and later important extensions by Zerilli and Moncrief \cite{Zerilli:1970se,Moncrief1974} for Schwarzschild (nonspinning) black holes. The theory was later extended by Teukolsky to handle perturbations of the Kerr metric~\cite{Teukolsky:1973ApJ}. Small perturbations of a field with spin-weight $s$ evolve on the Kerr geometry according to the Teukolsky master equation~\eqref{eq:teuk0}.

One important astrophysical application of black hole perturbation theory is to numerically simulate extreme mass ratio inspiral (EMRI) systems. An EMRI is comprised of a small mass--$q$ compact object orbiting a large mass--$M$ black hole, where $q \ll M$. EMRI systems emit gravitational radiation at low frequencies and are a key target of the upcoming Laser Interferometer Space Antenna (LISA) observatory~\cite{AmaroSeoane:2017las,Amaro-Seoane_2007,2010arXiv1009.1402A}. A standard method for studying EMRIs uses the perturbation theory of Kerr black holes in an approximation which treats the smaller compact object as point--like and structureless. When a rotating black hole is perturbed by a small, compact object the Teukolsky equation features a source term proportional to a Dirac delta distribution, $\delta(x)$, and possibly its derivatives. 

Many numerical methods have been developed for the Teukolsky equation. Due to the separability of the Teukolsky equation in the frequency domain, frequency-domain solvers are one popular class of methods that have been extensively developed~\cite{fujita2005new,finn2000gravitational,glampedakis2002approximating,drasco2006gravitational,macedo2022hyperboloidal} and work especially well for sources that have a discrete frequency spectrum or approximately so like with adiabatically-driven inspiral.

Time-domain solvers~\cite{lopez2003perturbative,khanna2004teukolsky,burko2007accurate,sundararajan2007towards,sundararajan2008towards,sundararajan2010binary,zenginouglu2011null,burko2016gravitational,field2023gpu,field2022discontinuous,sopuerta2006finite,racz2011numerical,harms2013numerical,csukas2019numerical}, on the other hand, can generally handle a broader range of problems including high-eccentricity orbits and inspiral-merger-ringdown orbital regimes. Time-domain solvers are especially efficient when the source has a very broad or continuous Fourier spectrum. However, because the Teukolsky equation is not separable in the time-domain, most method development has focused on $2+1$ solvers: after expanding the field in angular modes $\exp(-\mathrm{i} m \phi)$ (the m-modes do separate) we are left with a differential equation with two spatial dimensions and time. The first $2+1$, time-domain solver was based on a Lax-Wendroff scheme~\cite{krivan1997dynamics}. Since then, various $2+1$ time-domain Teukolsky solvers have been developed based around pseudo-spectral~\cite{harms2014new} and weighted essentially non-oscillatory (WENO)~\cite{field2023gpu} methods. In particular, pseudo-spectral solvers are particularly well suited for smooth solutions while WENO methods are able to handle problems with sharp features such as those that arise in the computation of the Aretakis charge~\cite{Aretakis_2013,Angelopoulos:2018yvt,Burko:2019qqx}. 

Despite the progress on time-domain methods for the source-free Teukolsky equation, particle-driven perturbations of the Kerr geometry remain challenging. To solve the distributionally-sourced Eq.~\eqref{eq:teuk0}, various regularized  numerical approaches~\cite{engquist2005discretization,tornberg2004numerical,sundararajan2007towards,sundararajan2008towards} have been proposed. To our knowledge, all current treatments approximate the Dirac delta as a narrow Gaussian or discrete representations over an extended range of grid points~\cite{sundararajan2007towards,sundararajan2008towards,harms2014new}. In both cases, these methods introduce small length scales (e.g., when the approximation is a narrow Gaussian), which can be a source of systematic error and will impose smaller timestep restrictions when evolved with an explicit time-stepping method. We note that an alternative effective-source approach can also be applied to this problem by allowing for analytic modifications to the source term~\cite{thornburg2017scalar,diener2012self,warburton2014applying}.

These problems have a clean solution for linear wave equations of one spatial variable. In previous work on the Regge-Wheeler-Zerilli and scalar wave equations, it was shown that a suitably modified multi-domain pseudo-spectral~\cite{canizares2010pseudospectral,canizares2009simulations} or discontinuous Galerkin (DG)~\cite{field2009discontinuous,field2022discontinuous} method can exactly treat the Dirac delta distribution. Both methods (i) provide spectral accuracy even at its location, (ii) are especially well-suited for smooth solutions, which is the case away from the Dirac delta distribution. The main complication in applying the techniques of Ref.~\cite{field2009discontinuous,field2022discontinuous} is that the Teukolsky equation is typically written in either $3+1$ or $2+1$ form. However, over the past few years, multiple works have considered expanding the solution in (spin-weighted) spherical harmonics, leading to a coupled system of $1+1$ wave equations. To our knowledge, the benefits of this approach were first sketched out in~\cite{nunez2010one,SteinThesis,dolan2013superradiant}, and the relevant equations written out in explicit form (in Boyer-Lindquist coordinates) to study source-free scalar~\cite{dolan2013superradiant} and gravitational~\cite{barack2017time} perturbations. And very recently this (source-free) system was also written in hyperboloidal coordinates and numerically solved with a symmetric integration method~\cite{markakis2023symmetric}.

In this paper, we describe a discontinuous Galerkin (DG) method~\cite{reed1973triangular,Hesthaven2008,cockburn2000development,cockburn2001devising,cockburn1998runge} for solving the distributionally-sourced, $s=0$ Teukolsky equation describing scalar waves on Kerr. The main contribution of our work is to extend the numerical flux construction of Refs.~\cite{fan2008generalized,field2009discontinuous,field2022discontinuous} to the coupled $1+1$ system, Eq. ~\eqref{eq:teuk0-1p1-first-layers}. Our DG method discretizes this coupled system written in fully first-order form and achieves global spectral accuracy everywhere in the computational domain including at the particle's location. Furthermore, at future null infinity (where the far-field signal is recorded) our method is super-convergent, thereby allowing for extremely accurate waveforms for a comparatively lower-resolution numerical grid. As a by-product of our work, we also consider two systems that arise from different choices in the first-order reduction variables finding that certain choices are numerically problematic.

This paper is organized as follows. Section \ref{Sec:Evolution} derives the $1+1$ system in fully first-order form and discusses its hyperbolicity. The outer boundary, where we need to supply radiation outgoing boundary conditions and extract the far-field waveform, is handled using the method of hyperboloidal layers. The nodal discontinuous Galerkin method, and its generalization for incorporating Dirac delta distributions, are summarized in Sec.~\ref{Sec:NumericalScheme}. Section~\ref{Sec:Experiments} documents several experiments testing our method, including convergence tests against exact solutions in special cases where they are known, energy luminosities for circular orbits, and late-time tail behavior of the scalar field. Several appendices collect technical results and report on an alternative (and numerically problematic) formulation of the system. Appendix~\ref{app:system_hyperbolicity} shows our first-order system is symmetric hyperbolic and well-posed.

\section{Evolution equations}
\label{Sec:Evolution}

\subsection{Teukolsky equation as a coupled 1+1D system}
\label{Sec:TeukEq1p1d}

The Teukolsky master equation describes scalar, vector, and tensor field perturbations in the space-time of
Kerr black holes~\cite{Teukolsky:1973ApJ}. In Boyer-Lindquist coordinates, this equation takes the form
\begin{eqnarray}
\label{eq:teuk0}
&&
-\left[\frac{(r^2 + a^2)^2 }{\Delta}-a^2\sin^2\theta\right]
         \partial_{tt}\Psi
-\frac{4 M a r}{\Delta}
         \partial_{t\phi}\Psi \nonumber \\
&&- 2s\left[r-\frac{M(r^2-a^2)}{\Delta}+ia\cos\theta\right]
         \partial_t\Psi\nonumber\\  
&&
+\,\Delta^{-s}\partial_r\left(\Delta^{s+1}\partial_r\Psi\right)
+\frac{1}{\sin\theta}\partial_\theta
\left(\sin\theta\partial_\theta\Psi\right)+\nonumber\\
&& \left[\frac{1}{\sin^2\theta}-\frac{a^2}{\Delta}\right] 
\partial_{\phi\phi}\Psi +\, 2s \left[\frac{a (r-M)}{\Delta} 
+ \frac{i \cos\theta}{\sin^2\theta}\right] \partial_\phi\Psi  \nonumber\\
&&- \left(s^2 \cot^2\theta - s \right) \Psi = -4 \pi \left(r^2 + a^2 \cos^2\theta \right) T  ,
\end{eqnarray}
where $M$ is the mass of the Kerr black hole, $a$ its angular momentum per unit mass, $\Delta = r^2 - 2 M r + a^2$,  
$s$ is the spin weight of the field, and $T$ is a source term. The $s = 0$ version of this equation describes scalar fields,
and it is within this simpler setting that we will develop our methods.

Most time-domain numerical solvers do not directly discretize Eq.~\eqref{eq:teuk0}. Instead, due to the axisymmetry of the Kerr spacetime, a set of decoupled 2+1D equations can be derived by separating out the field's azimuthal dependence.
However, when the source term describes a point particle, both the original 3+1D and the 2+1D systems are numerically challenging. 
For example, for the 3+1D system, if $T \propto \delta(r - r_p) \delta(\theta - \theta_p) \delta(\phi - \phi_p)$ then 
the solution is singular near $(r_p, \theta_p, \phi_p)$.
Moreover, the methods of~\cite{fan2008generalized,field2009discontinuous}, 
which exactly models the delta distribution as a modification 
to the numerical flux, are no longer directly applicable in 3+1D. 

To overcome this issue and to allow for exact treatment of the Dirac distribution, we follow Refs.~\cite{SteinThesis,dolan2013superradiant,barack2017time} and derive a 1+1D coupled system of equations obeyed by the scalar field. We first expand the solution
\begin{align}
\label{eq:anstaz}
\Psi(t,r,\theta,\phi) = \sum_{\ell=0}^{\infty} \sum_{m=-\ell}^{\ell} \Psi_{\ell m}(t,r) Y_{\ell m}(\theta, \phi) \,,
\end{align}
in terms of the ordinary scalar spherical harmonics $Y_{\ell m}(\theta, \phi)$. We follow the same definition and conventions as Ref.~\cite{Thorne:1980ru} and, in particular, the harmonics are orthonormal when integrated over the sphere. 
Substituting Eq.~\eqref{eq:anstaz} into Eq.~\eqref{eq:teuk0} and using well-known properties of spherical harmonics we arrive at
\begin{align}
\label{eq:teuk0-1p1-v1}
&\sum_{\ell, m}\left[-\left(r^{2}+a^{2}\right)^{2}\ddot{\Psi}_{\ell m} 
-4\mathrm{i}mMar\dot{\Psi}_{\ell m}
+\Delta\partial_{r}\left(\Delta\partial_{r}\Psi_{\ell m}\right)\right.\nonumber \\
&\left.+\left(a^{2}m^{2}-\ell(\ell+1)\Delta\right)\Psi_{\ell m} +\Delta a^{2}\sin^{2}\theta\ddot{\Psi}_{\ell m} \right]Y_{\ell m}\nonumber \\
& = -4 \pi \Delta \left(r^2 + a^2 \cos^2\theta \right) T \,.
\end{align}
Here we use an over-dot to denote $\partial / \partial t$ differentiation.
The term $\sin^{2}\theta Y_{\ell m}$ is responsible for mode coupling, and we use relations from Ref.~\cite{bailey_1933}
to rewrite it as
\begin{align}
\sin^{2}\theta Y_{\ell m} & = c^{L}_{\ell m}Y_{L m} \nonumber \\
\label{eq:coupling}
 &=c^{\ell-2}_{\ell m}Y_{\ell-2, m} + c^{\ell}_{\ell m} Y_{\ell m} + c^{\ell+2}_{\ell m} Y_{\ell+2, m} 
 \,,
\end{align}
where, for brevity, we will use Einstein summation notation over the repeated ``$L$'' index.
Appendix~\ref{app:coupling} provides the calculations needed to find these coupling coefficients. 
Multiplying by $\overline{Y}_{\ell' m'}$ and integrating over the sphere, we arrive at 
\begin{align}
\label{eq:teuk0-1p1-v2}
& -\left(r^{2}+a^{2}\right)^{2}\ddot{\Psi}_{\ell m}  -4\mathrm{i}mMar\dot{\Psi}_{\ell m}
+\Delta\partial_{r}\left(\Delta\partial_{r}\Psi_{\ell m}\right) \nonumber \\
& +\left(a^{2}m^{2}-\ell(\ell+1)\Delta\right)\Psi_{\ell m}
+\Delta a^{2}  c^{\ell}_{L m}\ddot{\Psi}_{L m} 
 = \hat{g}_{\ell m}(t,r) \,,
\end{align}
where the overline denotes complex conjugation, 
$ c^{\ell}_{L m}\ddot{\Psi}_{L m}  =  c^{\ell}_{(\ell-2) m}\ddot{\Psi}_{(\ell-2) m}
+  c^{\ell}_{\ell m}\ddot{\Psi}_{\ell m} + c^{\ell}_{(\ell+2) m}\ddot{\Psi}_{(\ell+2) m} $,
and
\begin{align}
\hat{g}_{\ell m}(t,r) =  -4 \pi \Delta \int  \left(r^2 + a^2 \cos^2\theta \right) \overline{Y}_{\ell m} T d\Omega \,,
\end{align}
is the source term.

We now factor out the large-$r$ behavior of $\Psi_{\ell m}$,
\begin{align}
\label{eq:1overR-subs}
    \Psi_{\ell m} = \frac{\psi_{\ell m}}{\sqrt{r^{2}+a^{2}}} \,,
\end{align}
and to map out the background spacetime, we introduce the tortoise coordinate, $r_{*}$, defined by
\begin{align}
\label{eq:BLtrnsf}
 d r_{*}=\frac{r^{2}+a^{2}}{\Delta}d r \,.
\end{align}
In terms of $\psi_{\ell m}$ and $r_{*}$, Eq.~\eqref{eq:teuk0-1p1-v2} becomes
\begin{align}
\label{eq:teuk0-1p1-v3}
& - \ddot{\psi}_{\ell m} + \partial_{r_*}\partial_{r_*} \psi_{\ell m} + \frac{\Delta a^2}{(r^2+a^2)^2}  c^{\ell}_{L m} \ddot{\psi}_{L m} \nonumber \\
&-\frac{4\mathrm{i}mMar}{(r^2+a^2)^2}\dot{\psi}_{\ell m} + V_{\ell m} (r)\psi_{\ell m} 
= g_{\ell m}(t,r_*) \,,
\end{align}
where $\psi = \psi(t,r_{*})$, $g_{\ell m} = \hat{g}_{\ell m} / (r^2+a^2)^{\frac{3}{2}}$ and,
\begin{align}
V_{\ell m}(r&)= \left[\frac{3r^{2}\Delta^{2}}{\left(r^{2}+a^{2}\right)^{4}}-\frac{2r\Delta (r-M)}{\left(r^{2}+a^{2}\right)^{3}}\right.\nonumber \\
 & \left.-\frac{\Delta^{2}}{\left(r^{2}+a^{2}\right)^{3}}+\frac{a^{2}m^{2}}{\left(r^{2}+a^{2}\right)^{2}}-\frac{\ell(\ell+1)\Delta}{\left(r^{2}+a^{2}\right)^{2}}\right] \label{eq:potential} \,.
\end{align}
The resulting differential equation's principle part is particularly simple~\cite{bardeen1972rotating} and
reduces to the ordinary wave operator when $a=0$.
Eq.~\eqref{eq:teuk0-1p1-v3} is the one we numerically solve after 
carrying out a reduction to fully first-order form
and transforming it to hyperboloidal coordinates; refer to Sec.~\ref{Sec:First-Order} and Sec.~\ref{Sec:Layers}.

\subsection{Distributional source term due to a scalar-charged particle in circular orbit}
\label{Sec:SourceTerm}

In this subsection, we provide expressions for the sourcing functions, $g_{\ell m}(t)$, that arise from projecting the 
scalar charge density, $T$, onto the spherical harmonics. Following the standard setup, we assume 
our problem arises from a small ``particle'' of mass $q$ orbiting a large-mass $M$ Kerr blackhole, where the mass ratio
satisfies $q/M \ll 1$. Ignoring radiation-reaction effects and specializing to circular and equatorial geodesic orbits, the
particle's scalar charge density is given by~\cite{warburton2010self}
\begin{align}
T = \frac{q}{r_p^2 u^t} \delta(r-r_p) \delta(\theta-\theta_p) \delta(\phi - \phi_p(t) ) \,,
\end{align}
where the constant $r_p$ denotes the particle's radial position, $\theta_p=\pi/2$ denotes the particle's polar angle, 
$u^t$ is the t-component of the particle's four-velocity, and 
$\phi_p(t)$ is the particle's angular location. Expressions for 
\begin{align}
u^t =  g^{t\phi} {\cal L} - g^{tt} {\cal E} \,, \qquad \phi_p(t) = \frac{v^3}{M(1+\tilde{a}v^3)} t\,,
\end{align}
are readily given in, for example, Refs.~\cite{warburton2010self,hughes2000evolution}. Here $v=\sqrt{M/r_p}$, $\tilde{a} = a/M$, and 
\begin{align}
{\cal E} = \frac{1 - 2v^2 + \tilde{a}v^3}{\sqrt{1 - 3v^2 +2\tilde{a}v^3}} \,, \qquad 
{\cal L} = r_p v\frac{1 - 2\tilde{a}v^3 + \tilde{a}^2v^4}{\sqrt{1 - 3v^2 +2\tilde{a}v^3}} \,,
\end{align}
are, respectively, the particle's energy and angular momentum. The background Kerr metric is denoted by $g_{\mu \nu}$.

With these preliminaries in place, for motion in the orbital plane and viewing $g_{\ell m}$ as a function of $r_*$, we now compute 
\begin{align}
\label{eq:source_glm2}
g_{\ell m}(t,r_*) & =
\frac{-4 \pi q}{u^t \left( r_p^2 + a^2 \right)^{1/2}} \overline{Y_{\ell m}}\left(\frac{\pi}{2},\phi_p(t)\right) \delta(r_{*} - r_{*,p}) \,,
\end{align}
where we have used
$\delta(r - r_p) = \left[(r_p^2 + a^2) / \Delta_p \right] \delta(r_{*} - r_{*,p})$ to transform the Dirac distribution into $r_*$ coordinates.

\subsection{Reduction to first-order form}
\label{Sec:First-Order}

For wave-like problems, 
the discontinuous Galerkin method we will introduce in Sec.~\ref{Sec:NumericalScheme} is most readily applicable
to systems in fully first-order form. Towards this end, we introduce two new variables, 
\begin{align} \label{eq:first_order_aux_vars_good}
 & \pi_{\ell m}=-\frac{\partial\psi_{\ell m}}{\partial t} \,, \qquad \phi_{\ell m}=\frac{\partial\psi_{\ell m}}{\partial r_*}\,.
\end{align}
The following first-order system corresponds to the original second-order wave equation~\eqref{eq:teuk0-1p1-v3}:
\begin{subequations}\label{eq:teuk0-1p1-first}
\begin{align}
\dot{\psi}_{\ell m} &=-\pi_{\ell m} \,, \\
\dot{\pi}_{\ell m} 
- f c^{\ell}_{Lm} \dot{\pi}_{L m} 
 & = - \partial_{r_*} \phi_{\ell m} - \mu\pi_{\ell m} \\  
 & \hspace{12pt}
- V_{\ell m} (r)\psi_{\ell m} + g_{\ell m}(t,r) \,, \nonumber \\
\dot{\phi}_{\ell m}  &= - \partial_{r_*}  \pi_{\ell m} \,.
\end{align}
\end{subequations}
In these expressions and below, we will use
\begin{align} \label{eq:def_mu_f}
f = \frac{\Delta  a^{2}} {\left(r^{2}+a^{2}\right)^{2}} \,, \qquad \mu =  \frac{4\mathrm{i}mMar} {\left(r^{2}+a^{2}\right)^{2}} \,,
\end{align}
for convenience. 
If $\psi_{\ell m}$ solves the first-order system~\eqref{eq:teuk0-1p1-first} it also solves the 
original second-order equation~\eqref{eq:teuk0-1p1-v3} provided the constraint $\phi_{\ell m}=\partial_{r_*} \psi_{\ell m}$ is satisfied. We note that alternative reduction variables could also be used. One such seemingly-reasonable choice ($\pi_{\ell m}=-\partial_{\tau} \psi_{\ell m}$ and $\phi_{\ell m}=\partial_{\rho} \psi_{\ell m}$ based on the hyperboloidal coordinates introduced in Sec.~\ref{Sec:Layers}) is considered in App.~\ref{app:alternative_system}, and in Sec.~\ref{Sec:Experiments} we demonstrate that this choice appears to be numerically problematic.

\subsection{Hyperboloidal layers}
\label{Sec:Layers}

The numerical simulation of wave phenomena on an open domain requires the specification of radiation boundary conditions 
and, in the case of gravitational-wave simulations, access to the waves that reach future null infinity.
We use the method of hyperboloidal layers~\cite{zenginouglu2011hyperboloidal,zenginouglu2011null} to solve both issues at the outermost physical boundary. 
We first introduce hyperboloidal coordinates, $(\rho, \tau)$, defined by
\begin{align} \label{eq:hyperboloidal_transformation}
r_{*}=\frac{\rho}{\Omega(\rho)} \,, \qquad \tau=t-h(r_*) \,,
\end{align}
that are related to the original $(r_*,t)$ coordinates through specification of the functions,
\begin{align} \label{eq:compression_and_height}
 & \Omega=1-\left(\frac{\rho-R}{s-R}\right)^{P}\Theta(\rho-R) \,, \quad  h=\frac{\rho}{\Omega}-\rho \,,
\end{align}
where $R$, $s$, and $P$ are to-be-set parameters and $\Theta$ is the Heaviside step function. 
To the left of the timelike surface defined by $\rho=R$, the coordinates are
the original $(r_*,t)$ ones. 
To the right of $\rho=R$, the coordinates smoothly connect the computational domain to future null infinity, which
is defined by $\rho=s$.
The width of the hyperboloidal layer is given by $s-R$.
We will choose the location of interface $R$ such that 
(i) the source term is always located to the left of the interface and (ii) for our multi-domain method, we collocate
the start of the layer at a subdomain interface. A sufficiently smooth coordinate transformation can be achieved by
setting the value of $P$, typically taken to be a positive integer~\cite{bernuzzi2011binary}. The value of $P$ could potentially impact the numerical scheme's order of convergence. For our multi-domain method, 
we anticipate that so long as $H$, given by Eq.~\eqref{eq:HypLayers}, is continuous at a subdomain's interface our DG method's convergence properties will not be impacted. This expectation is confirmed in Sec.~\ref{Sec:Exp_layers}.
Besides this one numerical experiment, we use $P=4$ throughout this paper.

In hyperboloidal coordinates, Eq.~\eqref{eq:teuk0-1p1-first} becomes~\footnote{When enacting this coordinate transformation
we do not let $\pi_{\ell m}$ and $\phi_{\ell m}$ transform. If we did, the resulting system would be 
the problematic one considered in App.~\ref{app:alternative_system}.} 
\begin{subequations}\label{eq:teuk0-1p1-first-layers}
\begin{align}
\dot{\psi}_{\ell m} & = -\pi_{\ell m} \,, \\
\dot{\pi}_{\ell m} 
- f c^{\ell}_{Lm} \dot{\pi}_{L m} 
- H \dot{\phi}_{\ell m}
 & =  - \left(1-H\right) \phi_{\ell m}^{\prime} \\
 & \hspace{-60pt} - \mu\pi_{\ell m} - V_{\ell m} (r)\psi_{\ell m} + g_{\ell m}(t,r) \,, \nonumber \\
\dot{\phi}_{\ell m} - H \dot{\pi}_{\ell m}   &= -  \left(1-H\right) \pi_{\ell m}^{\prime} \,,
\end{align}
\end{subequations}
where we now use an over-dot to denote $\partial / \partial \tau$ differentiation, 
a prime for differentiation by $\partial / \partial \rho$, and we note 
that $\partial / \partial \tau  = \partial / \partial t$.
Here  $H=\partial h / \partial r_* = 1 - \partial \rho / \partial r_*$ and, for later use, we note that
\begin{align} \label{eq:HypLayers}
H = 1 - \frac{\Omega^{2}}{\Omega - \rho \Omega^{\prime}} \,, \qquad
H^{\prime} = \frac{dH}{d\rho}=-\omega^{\prime}  \,,
\end{align}
where $\omega = \Omega^{2}/(\Omega - \rho \Omega^{\prime})$.
To the left of the layer, where $\rho < R$, we have $H=0$ and so both Eq.~\eqref{eq:teuk0-1p1-first}
and Eq.~\eqref{eq:teuk0-1p1-first-layers} are identical. One important fact of this 
coordinate transformation, which we will use later on, is that 
the outgoing characteristics obey $\tau - \rho = t - r_*$. 
While the coordinate transformation is singular at $\rho=s$, 
the coefficients of each term in Eq.~\eqref{eq:teuk0-1p1-first-layers}
are finite. This can be seen by noting that as $\rho \rightarrow s$ we have
$\left(1 - H\right) \sim \Omega^2 \sim r_*^{-2} \sim r^{-2}$ ~\cite{zenginouglu2011hyperboloidal,zenginouglu2011null}.

\subsection{Truncating to a finite number of modes}
\label{Sec:modes}

To achieve a finite number of equations, the expansion Eq.~\eqref{eq:anstaz} must be truncated to a 
finite value $\ell_{\rm max}$. Let $\Psi_{\rm full}$ be the infinite series~\eqref{eq:anstaz} and $\Psi_{\ell_{\rm max}}$ the truncated
expansion. Then the angular truncation error for $a=0$ (no mode mixing) is
\begin{align}
\Psi_{\rm full}  - \Psi_{\ell_{\rm max}} = \sum_{\ell={\ell_{\rm max}+1}}^{\infty} \sum_{m=-\ell}^{\ell} \Psi_{\ell m}(t,r) Y_{\ell m}(\theta, \phi) \,.
\end{align}
Integrating the residual over the sphere, we arrive at 
\begin{align}
\delta \Psi 
=  \sum_{\ell={\ell_{\rm max}+1}}^{\infty} \sum_{m=-\ell}^{\ell} \left| \Psi_{\ell m}(t,r) \right|^2 \nonumber \,,
\end{align}
where $\left| \cdot \right| $ is the complex modulus. 
Due to the expected decay of the coefficients $\Psi_{\ell m}(t,r)$, we can estimate the angular approximation error as 
$\delta \Psi \approx   \sum_{m=-\ell_{\rm max}}^{\ell_{\rm max}} \left| \Psi_{\ell m}(t,r)\right|^2$, which can be monitored throughout the simulation. 
For $a \neq 0$, $\ell$ modes will mix, but we can nonetheless use this estimation as a useful guide. While the angular approximation might seem like a limitation to our method, we point out that this error shows up under a different guise in any numerical scheme. For example, traditional 2+1 solvers that discretize the $\partial_{\theta}$ operator effectively set a resolution for the maximum resolvable $\ell$-mode. In all cases, one can estimate the error by increasing the angular grid resolution (for traditional 2+1 solvers) or $\ell_{\rm max}$ (our angular resolution error).

\subsection{Matrix form}
\label{Sec:matrix_form}

We now turn to writing Eq.~\eqref{eq:teuk0-1p1-first-layers} in matrix form, which will prove useful later on.
Let $\ell_{\rm max}$ be the largest value of $\ell$ we will consider in the computation, that is we
make an approximation $\psi_{\ell m }=0$ for $\ell > \ell_{\rm max}$. 
Noting that (i) the equation for 
odd and even $\ell$ modes decouple and (ii) each $m$ mode decouples, 
let us define a solution vector for the even-$\ell$ mode, $U_{\ell_{\rm max} m}^{\rm even} = \left[ \vec{\pi}_{\ell_{\rm max} m}^{\rm even}, \vec{\phi}_{\ell_{\rm max} m}^{\rm even} \right]$, where
\begin{align}
\vec{\pi}_{\ell_{\rm max} m}^{\rm even} =  \left[\pi_{\ell_{\rm min} m}^{\rm even}, \pi_{(\ell_{\rm min}+2)m}^{\rm even}, \dots, \pi_{\ell_{\rm max}m}^{\rm even}  \right]^{T} \,, \\
\vec{\phi}_{\ell_{\rm max} m}^{\rm even} = \left[\phi_{\ell_{\rm min}m}^{\rm even}, \phi_{(\ell_{\rm min}+2)m}^{\rm even}, \dots, \phi_{\ell_{\rm max}m}^{\rm even} \right]^{T} \,.
\end{align}
We have introduced $\ell_{\rm min}$ for the smallest value of $\ell$ given $m$ and the even/odd mode type. For example, if $m=0$ we have $\ell_{\rm min} = 0$ $(1)$ for the even (odd) modes, while if $m=2$, we have $\ell_{\rm min} = 2$ $(3)$ for the even (odd) modes. A similar set of notation is equally applicable to the odd-$\ell$ modes, leading to the system vector $U_{\ell_{\rm max} m}^{\rm odd}$.

The coupled system of equations~\eqref{eq:teuk0-1p1-first-layers}(b,c) take the form,
\begin{align} \label{eq:matrix_vector_system}
E\dot{U}+\hat{A}U^{\prime}+\hat{B}(U)+\hat{V}(\psi) = \hat{G}(t) \delta(r_{*} - r_{*,p}) \,,
\end{align}
where the components of $E$, $\hat{A}$, $\hat{B}$, and $\hat{V}$ can be read off from Eq.~\eqref{eq:teuk0-1p1-first-layers}.
The vector $\hat{G}(t)$ contains the coefficients of the distributional source term shown in Eq.~\eqref{eq:source_glm2}.
In this expression, we have dropped the $m$, $\ell_{\rm max}$, and $\{{\rm odd}, {\rm even}\}$ labels from both the matrices
and solution vector for clarity. The remaining set of equations, $\dot{\psi}_{\ell m} = -\pi_{\ell m}$, are trivially 
evolved along with the Eq.~\eqref{eq:matrix_vector_system}. Clearly, the size of these matrices change with $\ell_{\rm max}$ and $m$. For example, choosing $({\rm type},\ell_{\rm max},m)=({\rm even},8,0)$ will result in a $10$-by-$10$ system, while $({\rm type},\ell_{\rm max},m)=({\rm even},8,4)$ will yield a $6$-by-$6$ system. Inverting $E$, Eq.~\eqref{eq:matrix_vector_system} becomes 
\begin{align} \label{eq:matrix_vector_system2}
\dot{U}+AU^{\prime}+B(U)+V(\psi) = G(t) \delta(r_{*} - r_{*,p}) \,,
\end{align}
where we have defined $A=E^{-1} \hat{A}$, $B=E^{-1} \hat{B}$, $V=E^{-1} \hat{V}$, and $G=E^{-1} \hat{G}$.
We summarize the discretization of system, Eq. ~\eqref{eq:matrix_vector_system2} in Sec.~\ref{Sec:NumericalScheme}.

For concreteness, consider the even $\ell$-mode sector with $m=0$ and $\ell_{\rm max}=2$; App.~\ref{app:full_system} considers the most general system. 
Then we have $U=\left[\pi_{00}, \pi_{20}, \phi_{00}, \phi_{20},\right]^{T} $, and
\begin{align}
E=\begin{bmatrix}
1-fc_{00}^0 & -fc_{20}^0 & - H & 0\\
-fc_{00}^2 & 1-fc_{20}^2 & 0 & - H\\
-H & 0  & 1 & 0\\
0  & -H & 0 & 1
\end{bmatrix} \,,
\end{align}
and
\begin{align}
\hat{A}=\begin{bmatrix}
0 & 0 & (1-H) & 0\\
0 & 0 & 0      & (1-H)\\
(1-H) & 0 & 0 & 0\\
0 & (1-H) & 0 & 0
\end{bmatrix} \,,
\end{align}
with 
\begin{align}
\hat{V}&=\begin{bmatrix}V_{0}\psi_{00}, & V_{2}\psi_{20}, & 0, & 0\end{bmatrix}^{T} \,, \\
\hat{B}&=\begin{bmatrix} \mu \pi_{00}, & \mu \pi_{20}, & 0, & 0\end{bmatrix}^{T} \,. 
\end{align}
The components of the source vector are given by 
\begin{equation}
\hat{G}= \frac{-4 \pi q}{u^t \left( r_p^2 + a^2 \right)^{1/2}} \begin{bmatrix} 
\overline{Y_{0 0}}, &  \overline{Y_{2 0}}, & 0, & 0\end{bmatrix}^{T} \,,
\end{equation}
where the spherical harmonics are evaluated at $\theta = \pi/2$ and $\phi = \phi_p(t)$.

\subsection{System hyperbolicity}
\label{Sec:hyperbolicity}

We now consider the system's hyperbolicity, which has important consequences for the numerics and is directly used in the 
construction of the numerical flux. As shown in App.~\ref{app:full_system}, the system~\eqref{eq:matrix_vector_system} is (positive semi-definite) symmetric hyperbolic and well-posed. If we had instead used $\pi_{\ell m}=-\partial_{\tau} \psi_{\ell m}$ and $\phi_{\ell m}=\partial_{\rho} \psi_{\ell m}$ as our first-order reduction variables, the resulting system~\eqref{eq:teuk0-1p1-layers-first_alt} is not symmetric hyperbolic (although it is strongly hyperbolic) even when $a=0$. In principle, this isn't necessarily an issue, but we have found through extensive experimentation that our numerical scheme when applied to this alternative system shows a catastrophic loss of accuracy near $\rho = s$ (cf. Fig.~\ref{fig:convergence_at_scri} and Table~\ref{tab:fluxes_a0_vertical}). The observed non-convergent behavior may be due to the fact that the problem's boundary matrix~\footnote{In one spatial dimension, the right boundary matrix is simply $P = A(\rho = s)$. See Sec.~5 of Ref.~\cite{sarbach2012continuum}.}, is not invertible. This alternative system and these issues are discussed in App.~\ref{app:alternative_system}.

To study the system's wave structure, we consider the principle part of Eq.~\eqref{eq:matrix_vector_system2},
\begin{align} \label{eq:matrix_vector_system_PP1}
\dot{U}+AU^{\prime}+ \dots = 0 \,,
\end{align}
and diagonalize the matrix $A$
\begin{align} \label{eq:matrix_vector_system_PP2}
A(\rho) = T(\rho) \Lambda(\rho) T^{-1}(\rho) \,.
\end{align}
Let us assume $U$ is of length $L$, then $T$ is an $L$-by-$L$ matrix whose $i^{\rm th}$ column is the 
right eigenvector of $A$ corresponding to the eigenvalue $\lambda_i(\rho)$. The eigenvalues 
are real and correspond to the entries of the diagonal matrix $\Lambda = \mathrm{diag}(\lambda_1, \lambda_2, \dots, \lambda_L)$.
Furthermore, the eigenvalues satisfy $-1 \leq \lambda_i \leq 1$. Noting that $L$ is always even, half of the eigenvalues are $1$ and half of the eigenvalues are non-positive at $\rho < s$ and become exactly $0$ at $\rho=s$.

Assuming ``frozen'' matrix coefficients and letting $W=T^{-1}U$, the principle part of the system can be written as 
\begin{align} \label{eq:matrix_vector_system_diagonalized}
\dot{W}+\Lambda W^{\prime} = 0 \,.
\end{align}
Then, like the simple advection equation, 
$\lambda_i > 0$ corresponds to outgoing waves 
moving at a speed $\lambda_i$ while $\lambda_i < 0$ corresponds to incoming waves moving at a speed $\left|\lambda_i\right|$.
For later use in Sec.~\ref{sec:dg_delta}, 
we split the diagonal matrix as
$\Lambda = \Lambda^+ + \Lambda^-$, where $\Lambda^+$ ($\Lambda^-$) contains only positive (negative) eigenvalues. 
We then define the projection matrix $P^+$ as $\Lambda^+$ with non-zero entries replaced by $1$. Similarly,
let $P^-$ be $\Lambda^-$ with non-zero entries replaced by $1$. With this new notation, the non-zero components of $P^+W$ and $P^-W$
are, respectively, the right-moving and left-moving waves. These projection operators are used in the construction of the numerical flux given
Sec.~\ref{sec:dg_delta}.

\subsection{Boundary conditions}
\label{Sec:BCs}

Due to the hyperboloidal coordinate transformation, the wave speeds of the system, Eq. ~\eqref{eq:teuk0-1p1-first-layers} at the right physical boundary $\rho=s$ are either $1$ or $0$. 
This means $\rho=s$ is an outflow boundary, and hence no boundary conditions are needed. For the left boundary, we note that the system includes a mixture of incoming and outgoing characteristics and, furthermore, the potential is non-zero at the horizon when $m\neq0$ and $a\neq0$. We impose Sommerfeld boundary conditions setting the incoming characteristics to zero. This will lead to spurious reflection that can be mitigated by placing the left boundary at a sufficiently negative value of $\rho$ such that the left boundary is causally disconnected in the region of the computational domain that we care about. We note that this boundary issue can be partly solved by enacting so-called azimuthal transformations~\cite{krivan1996dynamics,PhysRevD.56.3395}, which we do not consider here except briefly in Sec.~\ref{sec:particle-waveforms_Kerr}.

\subsection{Energy flux}

One important application of the distributionally-sourced scalar wave equation is 
computing energy luminosities. Here, we provide a formula for this quantity in terms of 
our system variables. 

The radiative energy flux across an arbitrarily large ($r \rightarrow \infty$) spherical surface
will form the basis of some of our numerical experiments. 
The energy flux is given by~\cite{warburton2010self},
\begin{align}
\dot{E} = \frac{d E}{d t} = -\Delta \oint T_{tr} d \Omega \,,
\end{align}
where the stress-energy tensor
\begin{align}
T_{\alpha \beta} = \frac{1}{4\pi} \left( \partial_{\alpha} \Psi \partial_{\beta} \Psi - \frac{1}{2} g_{\alpha \beta} \partial^{\mu} \Psi \partial_{\mu} \Psi \right) \,,
\end{align}
is determined from derivatives of the scalar field.
Noting that the scalar field is real, we first write $T_{tr} = - \Psi_{,t} \overline{\Psi_{,r}} / (4 \pi)$, then
substitute
\begin{align}
\Psi(t,r,\theta,\phi) = \frac{1}{\sqrt{r^{2}+a^{2}}} \sum_{\ell=0}^{\infty} \sum_{m=-\ell}^{\ell} \psi_{\ell m}(t,r) Y_{\ell m}(\theta, \phi) \,,
\end{align}
to arrive at the individual multipole contributions to the total energy flux through the sphere at infinity,
\begin{align} \label{eq:energy_flux}
\dot{E} = - \frac{1}{4\pi} \sum_{\ell, m} \psi_{,t}^{\ell m}  \overline{\psi_{,r}^{\ell m}}
        =   \frac{1}{4\pi} \sum_{\ell, m} \left| \psi_{,t}^{\ell m} \right|^2 
        =   \frac{1}{4\pi} \sum_{\ell, m} \left| \pi^{\ell m} \right|^2  \,,
\end{align}
where we made use of the fact that
in the asymptotic limit ($r \rightarrow \infty$) the outgoing radiation condition 
implies $\psi_{,t} = - \psi_{,r}$. We will often directly compare the 
multipole contributions $\dot{E}_{\ell m} = \left| \pi^{\ell m} \right|^2 / (4 \pi)$.

\section{The discontinuous Galerkin Scheme}
\label{Sec:NumericalScheme}

Discontinuous Galerkin (DG) methods are especially well-suited for solving 
Eq.~\eqref{eq:matrix_vector_system2} and, more broadly, problems with Dirac delta distributions.
The DG method solves the weak form of the problem, a natural
setting for the delta distribution, and
the solution's non-smoothness can be 
``hidden'' at an interface between subdomains.
Our numerical scheme is based on the one described in Refs.~\cite{field2009discontinuous,field2022discontinuous}
for solving one-dimensional wave-like equations written in fully first-order form with source terms proportional to a Dirac delta distribution and its derivative(s). As such, we only summarize the key steps in the discretization and refer interested readers to those references for the details. After carrying out a spatial discretization using the nodal DG method, we integrate over time using the fourth-order Runge Kutta (RK4) scheme. 

\subsection{The source-free method for $G=0$} 

We partition the spatial domain into $K$ non-overlapping subdomains defined by the
partition points $\rho_0 < \rho_1 < \dots < \rho_K = s$
and denote $\mathsf{D}^j = [\rho_{j-1},\rho_j] $ as the $j^{\rm th}$ subdomain. In this one-dimensional setup,
the points $\{\rho_i\}_{i=1}^{K-1}$ locate the internal subdomain interfaces, and we require one of them to be 
the location of the Dirac distribution. Since we are solving a one-dimensional problem, the subdomains are line segments
and neighboring subdomains will intersect at a point. We note that non-ciruclar orbits with $r_p(t)$ can also be handled by using a time-dependent coordinate transformation, which has been worked out in simpler settings~\cite{field2009discontinuous}.

In each subdomain, 
all components of the solution vector $U$ and matrices $A$, $B$, and $V$ are expanded in a polynomial basis, which are taken
to be degree-$N$ Lagrange interpolating polynomials $\{ \ell_i(\rho) \}_{i=0}^N$ defined from Legendre-Gauss-Lobatto (LGL) nodes.
The time-dependent coefficients of this
expansion (e.g., on subdomain $j$: $\pi_{20}^j(\tau,\rho) = \sum_{i=0}^N \pi_{20}^i(\tau) \ell_i(\rho)$) 
are the unknowns we solve for. Products of terms arising in Eq.~\eqref{eq:matrix_vector_system2}
are represented through pointwise products of the interpolating polynomials at the LGL nodal points.

\begin{figure*}[!t]
  \centering
  %
  %
  \includegraphics[width=1.0\textwidth]{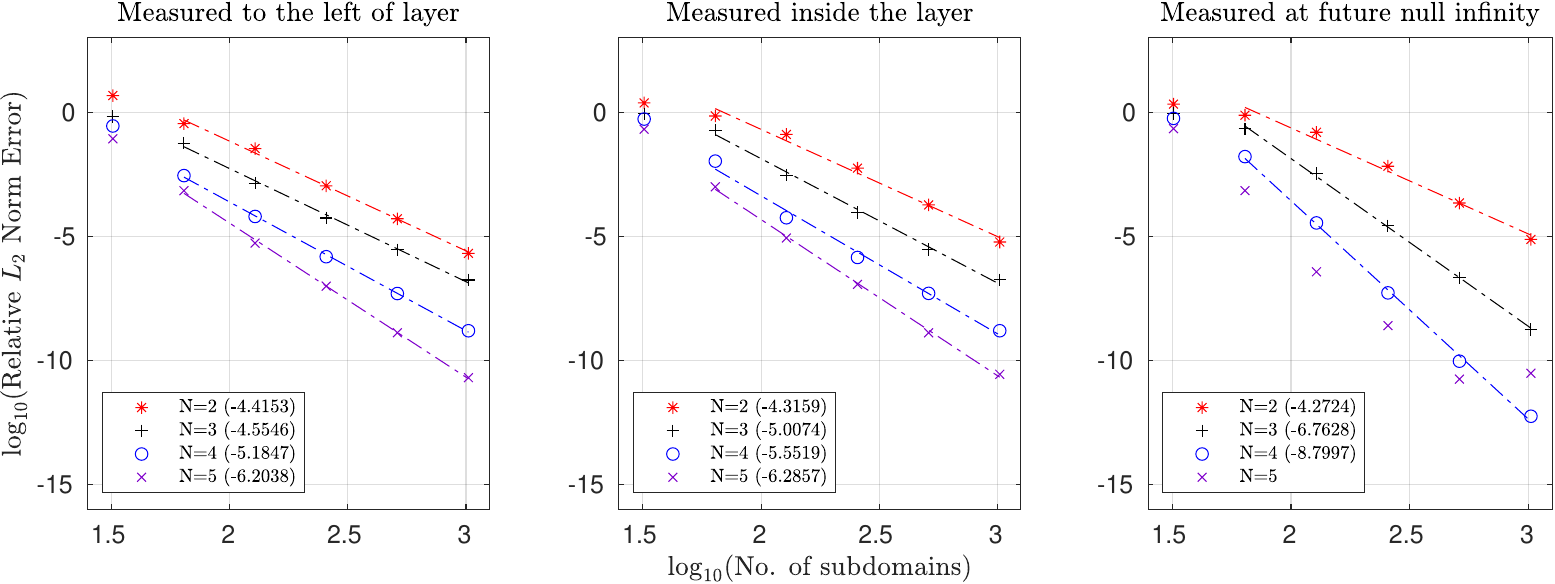}
        \caption{Convergence of the numerical solution to the exact solution for the $\ell=2$ wave equation using the hyperboloidal layer method. Errors are computed using a relative $L_2$-norm computation Eq. ~\eqref{eq:rel_L2}. We consider convergence by increasing the number of subdomains $K$ for approximating polynomial of degree $N=2$ (red asterisk), $N=3$ (black plus), $N=4$ (blue circle), and $N=5$ (purple cross). The error is measured at three locations on the spatial grid: before the start of the hyperboloidal layer (left panel), inside the hyperboloidal layer (middle panel), and at future null infinity given by $\rho=50$ (right panel). For a fixed value of $N$, the approximation error decreases with a power law $E \propto K^{-p}$. We empirically measure $p$ using a linear fit; our measured rates are shown in the legend. We find the convergence rate is similar to the ``standard'' one of $N+1$ before and inside the layers. At future null infinity, our scheme achieves superconvergence with a rate of nearly $2N+1$.
        }
  \label{fig:super_convergence}
\end{figure*} 

On each subdomain, we follow the standard DG recipe by requiring the residual 
\begin{equation} 
R_j(\tau,\rho) = 
\partial_{\tau}  U^j
+ A^j \partial_{\rho} U^j
+ B^j U^j
+ V^j \,,
\end{equation}
to satisfy 
\begin{equation} \label{eq:DG_1}
\int_{\mathsf{D}^j} R_j \ell_i^j d\rho  = 
\left[ \left(F^j - F^*\right) \ell_i^j \right]_{x_{j-1}}^{x^j}  \,,
\end{equation}
for all basis functions -- that is for all $i=0, 1, \dots, N$. Here we use a superscript ``$j$'' for vectors and matricies whose components have been approximated by Lagrange interpolating polynomials. Eq.~\eqref{eq:DG_1} features the physical flux vector, $F(U)=AU$ and the numerical flux $F^*$.  
The numerical flux is some yet-to-be-specified function $F^*(U^L, U^R)$, where $U^L$ and $U^R$
are, respectively, the left and right boundary values of the numerical solution
defined on the interface between neighboring subdomains. To build a stable and convergent DG scheme the numerical flux
must satisfy a few basic properties~\cite{Hesthaven2008} such as consistency $F^*(U, U) = F(U)$. While there 
are many reasonable choices for the numerical flux, because of its simplicity
and low cost-of-evaluation for our large, coupled system, we use the local Lax-Friedrichs (LLF) numerical
flux. At each interface, the LLF flux is computed as
\begin{equation} \label{eq:LLF}
F^*(U^L, U^R) = \frac{1}{2} \left[ F(U^L) + F(U^R)\right] -\frac{\lambda^{\rm LLF}}{2} \left( U^R- U^L \right) \,,
\end{equation}
where $\lambda^{\rm LLF}$ is the maximum eigenvalue of the flux Jacobian matrix $A$ evaluated at an interface; in our case $\lambda^{\rm LLF}=1$. The local Lax-Friedrichs flux is seen to be an average of the physical flux at the interface plus a dissipative part proportional to $\lambda^{\rm LLF}$, which is necessary to stabilize the scheme.

We note that the integrals appearing in Eq.~\eqref{eq:DG_1} can be pre-computed on a reference interval in terms of mass and differentiation matrices, leading to a coupled system of ordinary differential equations (see Eq.~(47) of Ref.~\cite{field2009discontinuous}) that can be integrated in time using RK4. A Courant-Friedrichs-Lewy condition restricts the largest stable timestep
associated with explicit numerical integration of Eq.~\eqref{eq:DG_1}. For a DG scheme, it is known that
$\Delta t_{\rm max} \propto C \Delta x_{\rm min}/\lambda_{\rm max}$, where $\lambda_{\rm max}=1$ is the largest
wavespeed and $\Delta x_{\rm min}$ is the smallest distance between neighboring Legendre-Gauss-Lobatto points on the physical grid.
The unknown scaling factor $C$ is typically of order unity. 
Finally, the global solution is taken to be a direct sum of the local solutions defined on each subdomain,
$U_h(\tau, \rho) = \bigoplus_{j=1}^K U_j(\tau, \rho)$.

\subsection{Inverting $\hat{E}$ and expressions at infinity}
\label{sec:E_at_infinity}

While not part of our DG scheme per se, an important numerical consideration is how to best invert $\hat{E}$. While one could numerically invert $\hat{E}$, we found that higher accuracies can be achieved through symbolic computations. That is, we compute $A=\hat{E}^{-1} \hat{A}$ symbolically, then export these expressions into code.

This symbolic approach works well for $\rho < s$. At $\rho=s$ the coordinate transformation~\eqref{eq:hyperboloidal_transformation} is singular yet, as remarked in Sec.~\ref{Sec:Layers}, the coefficients of the first-order differential equation~\eqref{eq:teuk0-1p1-first-layers} are well behaved. However, special care is needed at $\rho=s$. To better appreciate the issue at hand, consider setting $M=a=0$, then we end up with system~\eqref{eq:wave-1p1-first-layers}. The terms on the right-hand-side -- these are $(1-H)$ and $V$ -- behave like $r^{-2}$ as $r\rightarrow \infty$. Meanwhile, elements in the matrix $\hat{E}^{-1}$ behaves like $r^{2}$ as $r\rightarrow \infty$. Consequently, the coefficients and $\hat{E}^{-1}V$ need to be analytically supplied at $\rho=s$. This strategy continues to be applicable for the the general case (with non-zero values of $M$ and $a$) too, and some relevant expressions are provided in App.~\ref{app:System_at_infinity}. At grid points immediately to the left of $\rho=s$, the relevant coefficients in the partial differential operator might suffer from ill-conditioning (especially for $a\neq0$ and large values of $\ell_{\rm max}$); we experimented with variable precision computations but found no benefit.

\subsection{Including the Dirac delta distribution} 
\label{sec:dg_delta}

With a non-zero source term proportional to a Dirac delta distribution, the numerical flux evaluated at the interface $\rho=r_{*,p}$ will be modified through additional terms. The form of these new terms were derived in Eq.~(58) of Ref.~\cite{field2009discontinuous} and later extended in Ref.~\cite{field2022discontinuous}. Consider a Dirac delta distribution located at the interface between elements $\mathsf{D}^{k_p}$ and $\mathsf{D}^{k_{p+1}}$.
The basic idea is to note that the DG method is a weak formulation of the differential equation, where the numerical solution is made to satisfy Eq.~\eqref{eq:DG_1}. When we collocate the $\delta(r_{*} - r_{*,p})$ with a subdomain interface, we are faced with evaluating two relevant integrals for the subdomain to the left and right of $r_{*,p}$. We enforce the usual selection property of the Dirac distribution  
when integrated over the union $\mathsf{D}^{k_p} \cup \mathsf{D}^{k_{p+1}}$, yet we are free to choose how the individual integrals contribute to the total one. Following Ref.~\cite{field2009discontinuous}, we carry out a preferred splitting according to the wave dynamics of the problem. Applying the procedure of Ref.~\cite{field2009discontinuous} to our problem we arrive at
\begin{subequations}\label{eq:NumericalFluxModified}
\begin{align}
- \left(F^*\right)^{k_{p+1}}_{\text{left, modified}} & =
-\left(F^*\right)^{k_{p+1}}_{\text{left}} + T P^+T^{-1} G(t) \,,\\
-\left(F^*\right)^{k_{p}}_{\text{right, modified}} & =
-\left(F^*\right)^{k_{p}}_{\text{right}} - T P^-T^{-1} G(t) \, .
\end{align}
\end{subequations}
The negative signs in front of each $F^*$ instance can be understood by noting that 
the flux and source vectors are defined on different sides of the differential equation.
Compare, for example, Eq.~\eqref{eq:DG_1} and Eq.~\eqref{eq:matrix_vector_system2}.

\section{Numerical experiments}
\label{Sec:Experiments}

In this section, we perform several numerical experiments to benchmark our setup. In Sec.~\ref{Sec:Exp_layers}, we demonstrate spectral convergence and superconvergence of the algorithm when specialized to flatspace wave equation. In this simplified setting we can write down an exact analytical solution and perform unambiguous convergence tests that are impossible to do in Kerr. In Sec.~\ref{Sec:Tails} we verify known tail behavior at a finite distance and at future null infinity by numerically computing the solution's late-time power-law decay behavior and its dependence on $\ell$. Finally, in Sec.~\ref{Sec:particle-waveforms}, we numerically compute energy fluxes for circular orbits in Schwarzschild and Kerr geometries. We consider a variety of $(\ell, m)$ modes, orbital parameters, black hole spins, and also consider the impact of hyperboloidal coordinate transformation parameters and initial data on the quality of our results. The numerical experiments documented in Sec.~\ref{Sec:Tails} and \ref{Sec:particle-waveforms} set $M=1$.

\subsection{Verification of hyperboloidal layers: convergence and superconvergence}
\label{Sec:Exp_layers}

Our first experiment demonstrates the exponential convergence of our solver throughout the hyperboloidal layer. Upon setting $a=M=q=0$~\footnote{Clearly this setup does not correspond to an EMRI model. However, the resulting system~\eqref{eq:wave-1p1-first-layers} is an important special case for which exact solutions can be computed. This allows for unambiguous code tests that are impossible otherwise.}, the potential~\eqref{eq:potential} becomes $V_{\ell m} = - \ell (\ell +1)/r^2$ and $r_* =r$. And 
so Eq.~\eqref{eq:teuk0-1p1-v3} is just the ordinary flatspace wave equation:
\begin{align} \label{eq:flatspace}
\left[- \frac{\partial^2}{\partial t^2} + \frac{\partial^2}{\partial r^2}  -\frac{\ell(\ell+1)}{r^{2}} \right]\psi_{\ell m}  = 0 \,.
\end{align}
The first-order system in hyperboloidal coordinates is given from Eq.~\eqref{eq:teuk0-1p1-first-layers}:
\begin{subequations} \label{eq:wave-1p1-first-layers}
\begin{align}
& \dot{\psi}_{\ell m} = -\pi_{\ell m} \\
& \dot{\pi}_{\ell m} 
- H \dot{\phi}_{\ell m}
 =  -  \left(1-H\right) \phi_{\ell m}^{\prime}
+ \frac{\ell(\ell+1)}{r^2}\psi_{\ell m} 
 \\
& \dot{\phi}_{\ell m} - H \dot{\pi}_{\ell m}   = -  \left(1-H\right) \pi_{\ell m}^{\prime} \,.
\end{align}
\end{subequations}

Due to the simple setting, an exact closed-form solution can be found~\cite{field2015fast}. Setting $\ell=2$, the outgoing solution to Eq.~\eqref{eq:flatspace} is
\begin{align} 
\label{eq:outgoingEll2}
\psi(t,r) = f''(t-r) 
                + \frac{3}{r} f'(t-r) 
                + \frac{3}{r^2} f(t-r) \,,
\end{align}
where $f(u)$ is an underlying function of $u=t-r$, 
the prime indicating 
differentiation in argument, and we have
suppressed the harmonic indices.
Specification of the profile function,
\begin{align}
\label{eq:sinegauss}
f(u) = \sin \left[f_0 \left(u-u_0\right) \right] 
\mathrm{e}^{-c\left(u-u_0\right)^2} \,,
\end{align}
determines a purely outgoing multipole
solution~\cite{field2015fast}\footnote{Eq. (53) from Ref.\cite{field2015fast} (which is Eq.~(60) in arXiv version 1) has a typo.
This equation is corrected in arXiv version 2.}. 
Here $c$ characterizes the solution's spatial extent,
$f_0$ its ``central'' frequency, and $u_0$ its offset. 
A consequence of the hyperboloidal layer coordinate transformation, Eq. ~\eqref{eq:HypLayers},
is that $t-r = \tau - \rho$, which allows us to easily re-express the 
exact solution in $(\tau, \rho)$ coordinates. 

We solve Eq.~\eqref{eq:wave-1p1-first-layers} (with $\ell=2$) on
$\rho \in [1, 50]$, with the layer at $R=30$ and using $P=4$, set the final time $T = 50$,  
and choose a timestep $\Delta t = 10^{-5}$ sufficiently small enough such
that the Runge-Kutta's temporal error is well below the spatial discretization error.
Upon setting $f_0 = 2$, $c = 1$, and $u_0 = -10$, we 
take the initial data from the exact solution evaluated at $\tau=0$, which
is (numerically) zero at the left boundary and the start of the layer $\rho=R$.
At the left physical
boundary point we choose numerical fluxes that weakly 
enforce simple Sommerfeld (outgoing) boundary conditions,
which are sufficient for our purposes as the solution never impinges upon this boundary. 
Because of the property of the hyperboloidal layers, no boundary conditions are needed at $\rho=50$.  

We now check the convergence of our numerical solution 
against the exact solution when using hyperboloidal layers. 
It is well known that for a fixed value of polynomial degree $N$, the approximation
error (when computed in the $L_2$ norm over the spatial grid at a fixed time $T$) typically
decreases as a power law with an anticipated rate of $N+1$ for smooth solutions. 
We say this is ``anticipated'' because it follows from standard results of polynomial 
approximation theory and has nothing to do with the DG scheme, which can modify the expected rate
depending on the system and various method choices. 
Without using hyperboloidal layers our numerical scheme (for the ordinary wave equation) 
is identical to the one of Ref.~\cite{field2022discontinuous} where the expected
convergence rate of $N+1$ is 
demonstrated in Fig.~3 of that reference; similar results hold for non-linear problems~\cite{kidder2017spectre}.

We now show that our numerical approximation error continues to decay with a rate of $N+1$
when using the method of hyperboloidal layers. Furthermore, 
at certain locations of the grid we achieve {\em superconvergent} rates of $2N+1$.
We report the numerical error as a
relative $L_2$ norm
\begin{align} \label{eq:rel_L2}
E_{N,K}(\rho) = \frac{\int_{0}^T \left| \psi_{\rm num}(\tau,\rho) -  \psi_{\rm exact}(\tau,\rho) \right|^2 d\tau}{\int_{0}^T \left| \psi_{\rm exact}(\tau,\rho) \right|^2 d\tau}
\end{align}
taken over time at a fixed value of $\rho$.
The numerical solution $\psi_{\rm num}$ is computed using a degree $N$ approximating polynomial on a grid of $K$ subdomains. 

Fig.~\ref{fig:super_convergence} shows the scheme's rate of convergence as $N$ and $K$ are varied.
We measure the error before the layer at $\rho=15$ (left panel), inside the layer at $\rho=40$ (middle panel), and at future null infinity at $\rho=50$ (right panel). For the first two cases, these values of $\rho$ correspond to grid points that do not lie on the rightmost side of a 
subdomain. The scheme's rate of convergence is consistent with our $N+1$ expectation. By comparison, the rate of at future null infinity is found to be $2N+1$. This is a hallmark of a superconvergent DG scheme~\cite{adjerid2007discontinuous,cao2016superconvergence,ji2014superconvergent}, at which certain grid points (in this case, the outflow boundary is also a Radau nodal point) converge at a higher rate than that of other grid points. Noting that the transformation is parameterized by a positive integer $P$, we have numerically confirmed the same errors (and hence convergence rates) for $P=\{2,3,4,5,6,7\}$. This is to be expected for a multi-domain method such as ours whenever the layer starts at a subdomain interface.

Because the far-field waveform reaching gravitational-wave detectors is taken to be at future null infinity, superconvergence allows our scheme to obtain highly accurate waveforms for comparatively sparse computational grids. Furthermore, because the largest stable timestep scales as $N^{-2}$, our high-order waveform computation can be achieved with a larger step. For example, setting $N=4$ we achieve $9^{\rm th}$ order convergence. Without superconvergence this would only be possible with $N=8$, which translates into $4\times$ larger timesteps. We note that superconvergence is not a property of hyperboloidal layers, but rather our DG scheme. Yet the two work well together as $\rho=s$ is both a super-convergent point and exactly where we need to record the far-field waveform after applying the method of hyperboloidal layers. 

\begin{figure}
  \centering
  \includegraphics[width=0.45\textwidth]{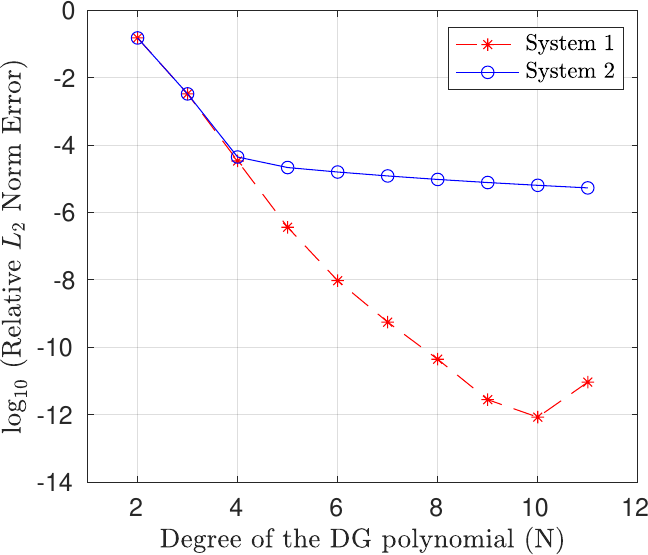}
        \caption{Spectral (exponential) convergence of the $\ell=2$ hyperboloidal layers experiment described in Sec.~\ref{Sec:Exp_layers}. Errors have been computed from $\psi_{\rm num}(\tau,\rho=50)$, which is the signal that reaches future null infinity $\mathscr{I^+}$. Convergence results are reported for system 1 (dashed-red line; our preferred first-order reduced system~\eqref{eq:wave-1p1-first-layers} that is symmetric hyperbolic) and system 2 (solid-blue line; an alternative first-order system~\eqref{eq:teuk0-1p1-layers-first_alt} that is discussed in App.~\ref{app:alternative_system}). For system 1, we see excellent convergence with $N$ until round-off error effects become noticeable around $10^{-12}$.  We find that our DG scheme solving system 2 is unable to obtain the expected accuracy that should be achievable by double precision. A complete understanding of this issue is still lacking; see App.~\ref{app:alternative_system}.} 
  \label{fig:convergence_at_scri}
\end{figure}

Finally, Fig.~\ref{fig:convergence_at_scri} demonstrates 
the spectral convergence of our method as applied to Eq.~\eqref{eq:wave-1p1-first-layers} (our preferred first-order reduced system; denoted ``System 1'' in the legend) and Eq.~\eqref{eq:teuk0-1p1-layers-first_alt} (denoted ``System 2'' in the legend). We observe the problematic ``System 2'' loses convergence much sooner than ``System 1'', in particular well before round-off error is expected to show up. While a complete understanding of this issue is still lacking, some observations are reported in App.~\ref{app:alternative_system}. For these experiments, we use a fixed number of $K=128$ subdomains and plot the error~\eqref{eq:rel_L2} as a function of $N$ and for each $N$. We have chosen a $\Delta t= 0.00001$ to ensure the temporal error is well below the spatial discretization error.

\subsection{Verification of Price tails}
\label{Sec:Tails}

The late-time decay behavior of scalar field perturbations around a Kerr black hole was first shown in Ref.~\cite{price1972nonspherical}. At late times, after the exponential decay of ringdown, the field decays as a power law $t^n$. In this subsection, we perform a series of experiments to test our code by comparing to results from the literature. In the legends of Figs.~\ref{fig:tails} and \ref{fig:tails_2}, the first value in the parentheses represents the expected behavior. We use two methods to extract a value of $n$ from our numerical data: (i) power law fit evaluated at $100,000$ (the second number in the parentheses of each legend) and (ii) each figure also includes an inset where the decay rate is computed from a finite-difference approximation to the logarithmic derivative $n=\partial_{\ln \tau} \ln\left( \psi_{\ell m}(\tau) \right)$. 

We provide initial data to the scalar field at some point away from the black hole event horizon but well before the hyperboloidal layer starts. The solution is then extracted as a time series at a location inside the layer and at $\mathscr{I}^{+}$. This also acts as an important test for the hyperboloidal layer method. 

Finally, we note that some results of this section have been compiled using the alternative first-order reduced system discussed in App.~\ref{app:alternative_system}; this shows that although the alternative system is problematic for certain studies (cf. Sec.~\ref{Sec:particle-waveforms} and Sec.~\ref{Sec:Exp_layers}) it can be used when high-accuracy is not required. 

\subsubsection{Price tails on Schwarzschild}
\label{Sec:Tails1}

\begin{figure*}[!t]
    \centering
    \includegraphics[width=0.95\columnwidth]{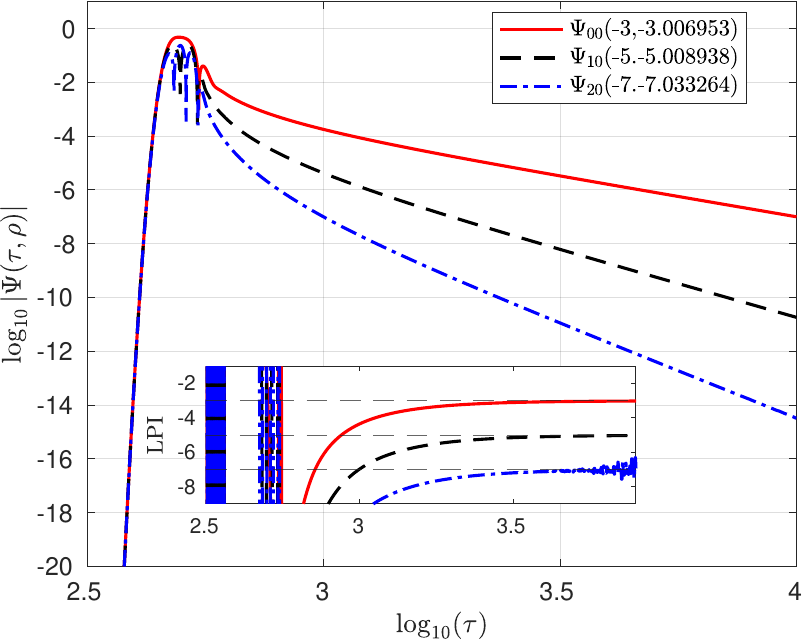}
    \includegraphics[width=0.95\columnwidth]{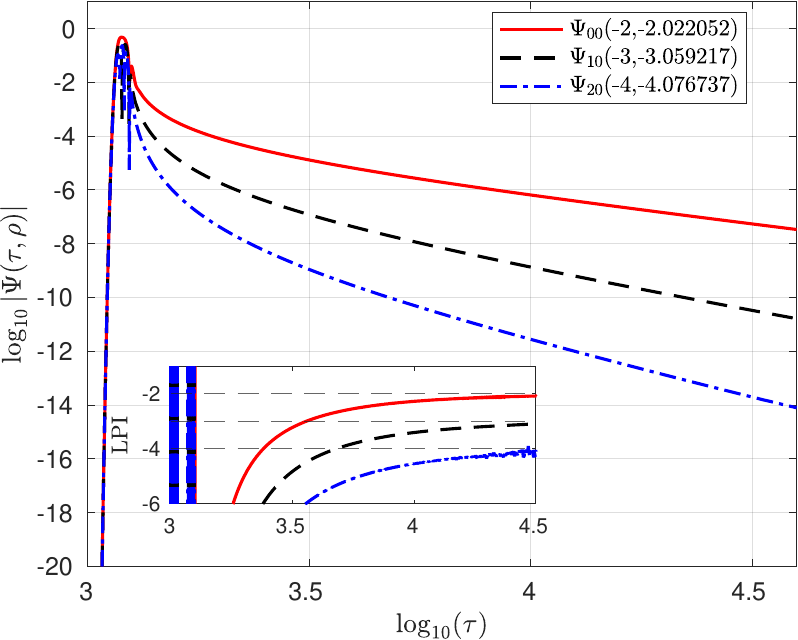}
    \caption{Late time tail behavior of a scalar field evolving on a Schwarzschild spacetime with non-zero momentum initial data, Eq.~\eqref{eq:non-zero-momentum-initial-data}. The 
    $\ell=0$ (solid red), $\ell=1$ (dashed black), and $\ell=2$ (dashed-dot blue) modes are shown. The inset figure shows the local power index (LPI) approaching the theoretically expected values shown as a dashed horizontal line. The parentheses values in the legend represent the expected tail rate and the numerical fit values.
    {\bf Left panel}: Late-time Schwarzschild tails for signals extracted at $r_{*}=500$ (inside the hyperboloidal layer) showing the expected $-2\ell-3$ power law behavior. {\bf Right panel}: The signal extracted at the right boundary which, from hyperboloidal layers compactification is the null infinity $\mathscr{I}^{+}$. We find the power-law behavior follows the expected $-\ell-2$ rate.}
    \label{fig:tails}
\end{figure*}

For our first test, we perturb the scalar field around the background Schwarzschild spacetime with non-zero momentum initial data,
\begin{align}
\psi_{\ell m} = 0 \,, \qquad  \pi_{\ell m} =  \frac{1}{\sqrt{2\pi10^{2}}}\exp \left( \frac{-(\rho - 30)^2}{2\pi10^{2}} \right)  \,,
\label{eq:non-zero-momentum-initial-data}
\end{align}
and where $\phi_{\ell m} = \partial_{r_*} \psi_{\ell m}$. We solve Eq.~\eqref{eq:teuk0-1p1-first-layers} (with $\ell=2$ and $M=1$) on $\rho \in [-200, 1200]$ with the layer at $R=150$. We use $K=400$ subdomains and on each subdomain the solution is approximated by a degree $N=10$ polynomial. The time step is set to $\Delta t = 0.0866$. In Fig.~\ref{fig:tails}, we show the solution recorded as a time series at $r_{*}=500$ (left panel) and at $\mathscr{I}^{+}$ (right panel). At late times we empirically measure the decay rate of solution and quote this value in the parentheses of the figure's label. In the inset figure, we also report the local power index and show the expected rate (as a horizontal line) based on Price's Law~\cite{price1972nonspherical}. Our measured tail decay rates are consistent with the theoretically known values at both a finite distance and $\mathscr{I}^{+}$~\cite{hod1999mode}.

\subsubsection{Price tails on Kerr}

Unlike the Schwarzschild case, a full understanding of Price tails in Kerr spacetime was not developed until relatively recently~\cite{GPP,TKT,burko-khanna09,burko2011late,zkb,burko2014mode} (the reader is referred to the detail therein). Consider the situation where $\ell'$ is the mode where we set the initial data (assuming non-zero momentum), and $\ell$ is the mode excited through coupling. As proposed in Ref.~\cite{zkb} and later shown in Ref.~\cite{burko2014mode}, except for the case in which $\ell'$ is the slowest decaying mode (for which case the decay rates are given by $-n=\ell'+\ell+3$), all other azimuthal modes decay along $r={\rm const}$ (approaching future timelike infinity $i^+$) according to $-n=\ell'+\ell+1$. Ref.~\cite{zkb} also found the decay rate along future null infinity $\mathscr{I^+}$, specifically, $-n^{\mathscr{I^+}}=\ell+2$ if $\ell\ge\ell'$, and $-n^{\mathscr{I^+}}=\ell$ if $\ell\le\ell'-2$.

\begin{figure}
    \centering
    \includegraphics[width=0.95\columnwidth]{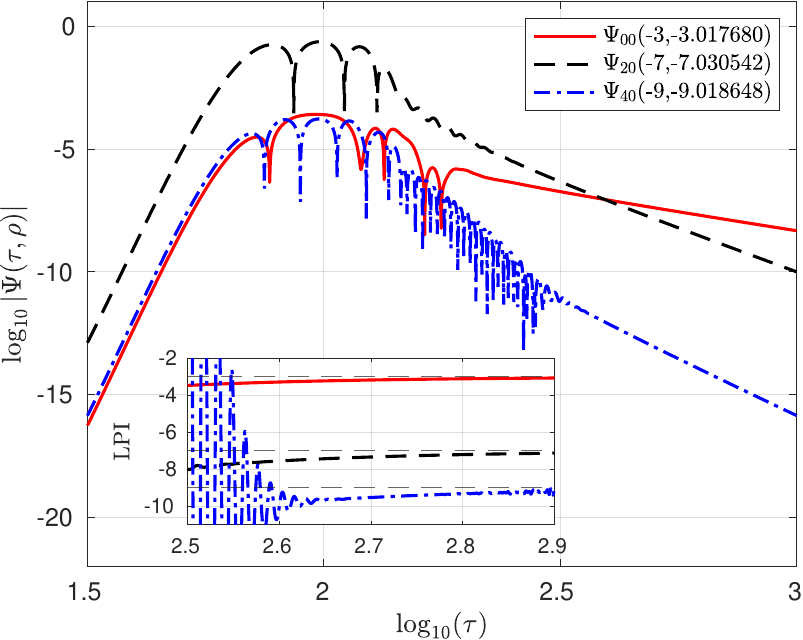}
    \caption{Late time tail behavior of a scalar field evolving on a Kerr spacetime with a spin of $a=0.99995$. The initial data is set as $\ensuremath{\psi_{2 0}=0}$, $\phi_{2 0} = 0$, and $\pi_{2 0}=\frac{1}{\sqrt{2\pi6^{2}}}\exp\left(\frac{-\left(\rho-25\right)^{2}}{2\pi6^{2}}\right)$, while the $(4,0$ and $(0,0)$ modes are 0. The solution is then extracted at $r_{*}=100$. The numbers in the parentheses show the expected rate vs the calculated decay rate of the perturbations.}
    \label{fig:tails_2}
\end{figure}

In Fig.~\ref{fig:tails_2} we repeat our Schwarzschild tails experiment but now setting $a=0.99995$. We solve on the domain $\rho \in [-2100,400]$ with the start of the hyperboloidal layer at $R=149.3500$. We use $K=400$ subdomains and on each subdomain the solution is approximated by a degree $N=15$ polynomial. The time step is set to $\Delta t = 0.0761$. To facilitate a direct comparison with previous work, we choose our initial data to match that of Ref.~\cite{burko2011late}. In particular, we provide a non-zero $(2,0)$ mode perturbation that, through mode coupling, excites the $(0,0)$ and $(4,0)$ modes. Technically all $m=0$, even $\ell$ modes will be excited, but we only consider nearest neighbor coupling for this experiment; see Sec.~\ref{Sec:modes}. Our measured rates are consistent with known results. 

As a final test, we extend this experimental setup (now using $N=10$, $K=400$, and  $\Delta t = 0.0436$) for a range of non-zero $(\ell',m)$ mode data and report the measured decay rate of the coupled $(\ell,m)$ modes. Table~\ref{tab:scattering_matrix} summarizes numerical experiments while using a spin value of $a=0.8$.

{\renewcommand{\arraystretch}{1.5}%
\begin{table*}[t]
\centering
\begin{tabular}{|c|c|c|c|c|c|c|}
\hline
\begin{tabular}[c]{@{}c@{}}Initial\\ $\ell',m$ mode\end{tabular}    & \begin{tabular}[c]{@{}c@{}}Projected\\ $\ell=0$\end{tabular}     & \begin{tabular}[c]{@{}c@{}}Projected\\ $\ell=1$\end{tabular}     & \begin{tabular}[c]{@{}c@{}}Projected\\ $\ell=2$\end{tabular}          & \begin{tabular}[c]{@{}c@{}}Projected\\ $\ell=3$\end{tabular}     & \begin{tabular}[c]{@{}c@{}}Projected\\ $\ell=4$\end{tabular}          & \begin{tabular}[c]{@{}c@{}}Projected\\ $\ell=5$\end{tabular}          \\ \hline
\begin{tabular}[c]{@{}c@{}}(0,0)\\ (1,1)\\ (2,0)\\ (2,2)\end{tabular} & \begin{tabular}[c]{@{}c@{}}-3.012\\ *\\ -3.010\\ *\end{tabular} & \begin{tabular}[c]{@{}c@{}}*\\ -5.008\\ *\\ *\end{tabular} & \begin{tabular}[c]{@{}c@{}}-5.027\\ *\\ -7.013\\ -7.016\end{tabular} & \begin{tabular}[c]{@{}c@{}}*\\ -7.016\\ *\\ *\end{tabular} & \begin{tabular}[c]{@{}c@{}}-7.449\\ *\\ -9.004\\ -9.049\end{tabular} & \begin{tabular}[c]{@{}c@{}}*\\ -9.710\\ *\\ *\end{tabular} \\
(3,1)                                                                 & *                                                               & -5.009                                                     & *                                                                    & -9.013                                                     & *                                                                  & -11.024                                                    \\
\begin{tabular}[c]{@{}c@{}}(4,0)\\ (4,2)\end{tabular}                 & \begin{tabular}[c]{@{}c@{}}-5.022\\ *\end{tabular}              & \begin{tabular}[c]{@{}c@{}}*\\ *\end{tabular}              & \begin{tabular}[c]{@{}c@{}}-7.010\\ -7.012\end{tabular}              & \begin{tabular}[c]{@{}c@{}}*\\ *\end{tabular}              & \begin{tabular}[c]{@{}c@{}}-11.255\\ -11.009\end{tabular}            & \begin{tabular}[c]{@{}c@{}}*\\ *\end{tabular}              \\ \hline
\end{tabular}
\caption{We reproduce the $(\ell',m)\rightarrow(\ell,m)$ scattering matrix from Ref.~\cite{burko2011late} where they verified the late time tail behavior of scalar fields in the Kerr background using quad precision. We set the spin on the primary black hole as $a=0.8$. The expected behavior is $\ell'+\ell+1$ if $\ell<\ell'$ and $\ell'+\ell+3$ if $\ell\geq\ell'$. We get the expected behavior with less than $1\%$ of error for the lower $\ell = \{0,1,2,3\}$ modes. Since the higher modes are very weakly coupled to the initial $\ell'$ mode, double-precision round-off errors prevent us from obtaining similar $1\%$ accuracy computations of the power-law behavior for some of the higher $\ell=\{4,5\}$ modes. More accurate results can be obtained with quad-precision floating-point computations, which we hope to consider in the future.}
    \label{tab:scattering_matrix}
\end{table*}

\subsection{Fluxes from a scalar-charged particle in circular orbit}
\label{Sec:particle-waveforms}

This subsection compares our circular orbit energy fluxes~\eqref{eq:energy_flux} 
to those obtained by other authors and codes. 

\subsubsection{Fluxes on Schwarzschild}
\label{sec:particle-waveforms_Schwarzschild}

We first consider the simpler Schwarzschild case, where $a=0$ and the harmonic 
modes decouple for all values of $(\ell,m)$. For our simulations, we have typically chosen 
$M = 1$, and we solve Eq.~\eqref{eq:teuk0-1p1-first-layers} with a distributional 
source term given by Eq.~\eqref{eq:source_glm2}. Our physical domain
$\rho \in [-100, 200]$ is partitioned into 200 subdomains
and on each subdomain we approximate the solution with a degree 
$N=10$ polynomial. We set the final time $T = 8000$, 
and the time step $\Delta t = 0.0018$. As is common practice,
we supply trivial (zero) initial
data. We note that supplying trivial initial
data is commonly done because (i) the exact initial data is unknown and 
(ii) the expectation is that over long enough times the impact of incorrect 
initial data will propagate away as so-called junk radiation~\cite{field2010persistent}.
Trivial initial data is inconsistent with the distributional forcing
terms~\eqref{eq:source_glm2}, so we smoothly turn on the source term with 
a ramp-up function. We take our ramp-up function to be 
Eq. C1 of Ref.~\cite{field2009discontinuous}
with parameter values $\tau = 400$  and $\delta = 0.00025$. 
To achieve high accuracies for some of the higher harmonic modes, we 
sometimes deviate from these default settings.

\begin{figure}
    \centering
    \includegraphics[width=0.48\textwidth]{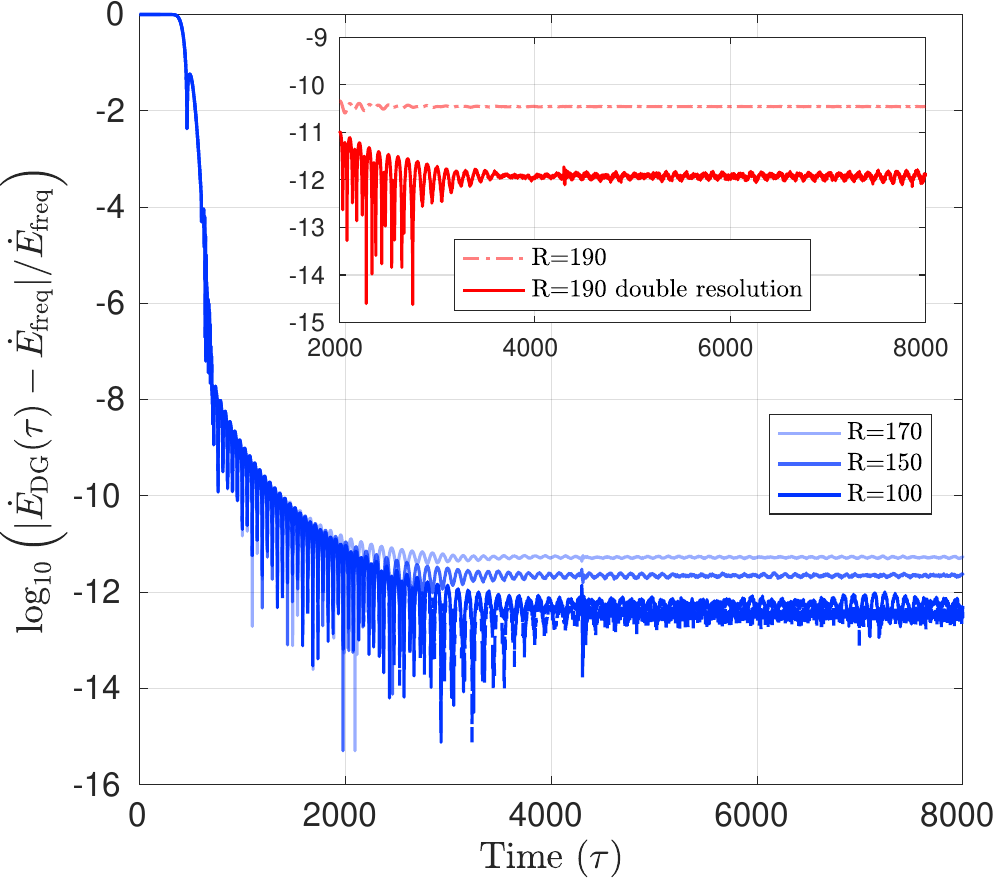}
\label{fig:Scalar_fluxes_convergence}
\caption{Relative error in the $(2,2)$-mode energy flux computed with our time-domain solver ($\dot{E}^{22}_{\rm DG}$) compared with a 
frequency-domain solver ($\dot{E}^{22}_{\rm freq}$). We find that increasing the hyperboloidal layer's width helps in achieving higher accuracy scalar fluxes in the $(2,2)$ mode. Here $R$ is the start of the layer, so larger values of $R$ correspond to narrower layers. We also find that spurious junk tails, due to inconsistent initial data, need sufficient time to decay away before high-accuracy measurements can be made. For this experiment, $N=9$, $K=100$ and $\Delta t= 0.010461$ were kept fixed for all values of $R$, and we set $K=200$ when using ``double resolution''.
}
\end{figure}

\begin{figure}
    \centering
    \includegraphics[width=0.48\textwidth]{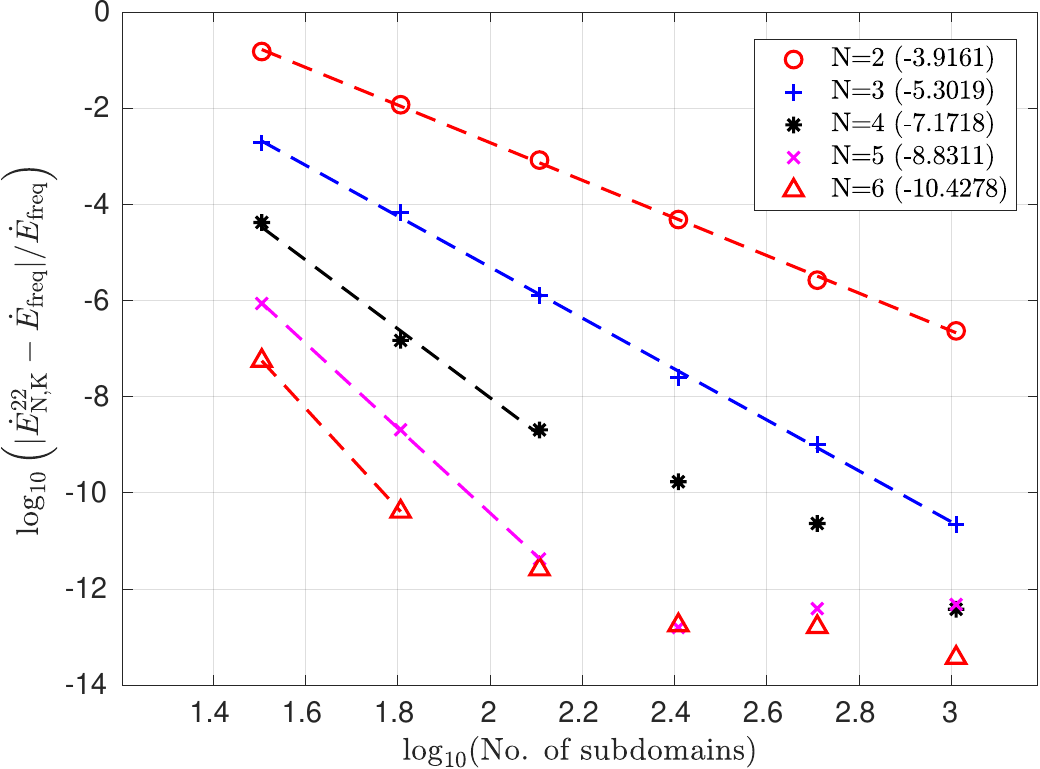}
\caption{Superconvergence of numerically computed energy fluxes from a particle in circular orbit at $r_{p}=10M$ on Schwarzschild. Relative errors in the $(2,2)$-mode energy flux are computed with our time-domain solver ($\dot{E}^{22}_{\rm N, K}$) and compared with a frequency-domain solver ($\dot{E}^{22}_{\rm freq} = 3.369977470603454 \times 10^{-6}$). We consider convergence by increasing the number of subdomains $K$ for approximating polynomial of degree $N=2$ (red circles), $N=3$ (blue plus), $N=4$ (black asterisk), $N=5$ (purple cross), and $N=6$ (red triangle). We find the convergence rate is between $2N-1$ to $2N$, which is faster than the ``standard'' one of $N+1$. Similar superconvergence was obtained for the ordinary wave equation using hyperboloidal layers; see Fig.~\ref{fig:super_convergence}.}
\label{fig:Scalar_fluxes_convergence_K}
\end{figure}

Before presenting our numerical results, we remark on potential sources of error.
With the numerical setup described above, the relative
error associated with our numerical solution before the start of the hyperboloidal layer 
is better than $10^{-10}$, which is sufficient for our purposes. At $\mathscr{I^+}$,
where the energy flux is computed, we encounter two 
additional sources of error: spurious junk due to trivial initial data and
setting the hyperboloidal layer's width. These two sources of (systematic) error
are quantified in Fig.~\ref{fig:Scalar_fluxes_convergence}. In this experiment, 
we consider a high-resolution numerical computation of the 
$(2,2)$-mode energy flux $\dot{E}^{22}_{\rm DG}(\tau)$ from 
a particle in circular orbit located at
$r_{*,p}=12.7725887222397$ (which corresponds to $r_p = 10$ in $r$ coordinates) and our domain is 
$\rho \in [r_{*,p}-200,r_{*,p}+200]$.
For a fixed location 
of the hyperboloidal layer's start ($\rho = R$), the figure
shows the agreement between our DG scheme and a frequency-domain solver~\cite{BHPToolkit,TeukFreqCode}
becomes better as we wait longer. Indeed, for each $R$ we plot the relative error computed as
\begin{align}
  \label{eq:rel_error_flux}
|\dot{E}^{\ell m}_{\rm DG}(\tau) - \dot{E}^{\ell m}_{\rm FR}| / |\dot{E}^{\ell m}_{\rm FR}| \,,
\end{align}
where $\dot{E}^{\ell m}_{\rm FR}$ is the energy flux computed with a frequency-domain solver and $\dot{E}^{\ell m}_{\rm DG}(\tau)$ is 
the time-domain DG solver's value. We see that at early times the solution is highly contaminated by spurious junk. Achieving a high-accuracy solution requires that we wait sufficiently long for the spurious ``junk tails'' to decay away. 
As noted in Ref.~\cite{field2010persistent}, because tails at $\mathscr{I^+}$ decay more slowly, one must wait even longer (as compared to finite-radius measurements) for these transients to die off. 

Fig.~\ref{fig:Scalar_fluxes_convergence} also shows that as 
$R \rightarrow s$ (ie the layer's width goes to zero) our solution quality degrades. We 
believe this is due to sharper gradients that stem from the compression function $\Omega$, 
which makes it harder for the numerical scheme to approximate the solution for a fixed grid resolution.
For example, setting $R=100$ (where the layer's width is $\approx 110$) we find the energy flux computation is accurate to a relative error on the order of $10^{-12}$. Keeping the grid resolution fixed, we see that the error in the flux computation increases as the layer's width shrinks. When $R=190$ (where the layer's width is $\approx 22$) the error is now $\approx 5 \times 10^{-11}$, yet doubling the number of subdomains improves the solution quality allowing for energy flux computations to again obtain relative errors on the order of $10^{-12}$. To achieve efficient numerical schemes it would be helpful to have criteria for setting the layer's width given the properties of the problem, but this is currently unavailable in the literature. 

Having properly accounted for these sources of error, Table \ref{tab:fluxes_a0_vertical} compares the energy flux (computed from high-resolution runs) to values computed using from a frequency-domain solver~\cite{TeukFreqCode} implemented within the Black Hole Perturbation Toolkit~\cite{BHPToolkit}. These frequency-domain results rely on the appropriate boundary value problems in the frequency domain and allows for a direct, non-trivial comparison between methods. Finally, Fig.~\ref{fig:Scalar_fluxes_convergence_K} shows the rate of convergence in our DG scheme's energy flux computation (at time $4000M$ after junk radiation has left the system), as $N$ and $K$ are varied while the timestep $\Delta t = 0.010460592$ is kept fixed. As the energy flux is computed at future null infinity we anticipate superconvergent rates similar to those observed for the flatspace wave equation; see the rightmost panel of Fig.~\ref{fig:super_convergence}. For DG methods the numerical error is expected to decrease exponentially fast, $E \propto K^{-p}$, where $p=N+1$ is the standard convergence rate. Fig.~\ref{fig:Scalar_fluxes_convergence_K} shows the rate of convergence to be much faster than $N+1$, and about 1 order of convergence less than what is shown in Fig.~\ref{fig:super_convergence}. This discrepancy can be understood by noting that the errors shown in Fig.~\ref{fig:super_convergence} are for $\Psi$, while for the energy flux computation, the errors are due to the numerical approximation of $\pi = -\partial_t \psi$, which is a derivative of $\Psi$. 

{\renewcommand{\arraystretch}{1.5}%
\begin{table}[t]
\begin{tabular}{cccc}
\hline
\multicolumn{1}{|c|}{($\ell$,m)} &
  \multicolumn{1}{c|}{Alg.} &
  \multicolumn{1}{c|}{Energy Flux ($\dot{E}^{\ell m}/q^2$)} &
  \multicolumn{1}{c|}{Relative Error} \\ \hline
\multicolumn{4}{c}{System 1} \\ \hline
\multicolumn{1}{|c|}{(2,2)} &
  \multicolumn{1}{c|}{\begin{tabular}[c]{@{}c@{}}FR\\ DG\end{tabular}} &
  \multicolumn{1}{c|}{\begin{tabular}[c]{@{}c@{}}$2.1676683889035\times 10^{-6}$\\ $2.1676683889071\times 10^{-6}$\end{tabular}} &
  \multicolumn{1}{c|}{$1.7\times 10^{-12}$} \\ \hline
\multicolumn{1}{|c|}{(4,2)} &
  \multicolumn{1}{c|}{\begin{tabular}[c]{@{}c@{}}FR\\ DG\end{tabular}} &
  \multicolumn{1}{c|}{\begin{tabular}[c]{@{}c@{}}$3.7609900151243\times 10^{-11}$\\ $3.7609900156162\times 10^{-11}$\end{tabular}} &
  \multicolumn{1}{c|}{$1.3\times 10^{-10}$} \\ \hline
\multicolumn{1}{|c|}{(5,3)} &
  \multicolumn{1}{c|}{\begin{tabular}[c]{@{}c@{}}FR\\ DG\end{tabular}} &
  \multicolumn{1}{c|}{\begin{tabular}[c]{@{}c@{}}$8.8539629089954\times 10^{-13}$\\ $8.8539629113568\times 10^{-13}$\end{tabular}} &
  \multicolumn{1}{c|}{$2.7\times 10^{-10}$} \\ \hline
\multicolumn{1}{|c|}{(9,7)} &
  \multicolumn{1}{c|}{\begin{tabular}[c]{@{}c@{}}FR\\ DG\end{tabular}} &
  \multicolumn{1}{c|}{\begin{tabular}[c]{@{}c@{}}$3.5707073944101\times 10^{-14}$\\ $3.5707073942955\times 10^{-14}$\end{tabular}} &
  \multicolumn{1}{c|}{$3.2\times 10^{-11}$} \\ \hline
\multicolumn{1}{|c|}{(15,15)} &
  \multicolumn{1}{c|}{\begin{tabular}[c]{@{}c@{}}FR\\ DG\end{tabular}} &
  \multicolumn{1}{c|}{\begin{tabular}[c]{@{}c@{}}$2.1814822732028\times 10^{-16}$\\ $2.1814822730317\times 10^{-16}$\end{tabular}} &
  \multicolumn{1}{c|}{$7.8\times 10^{-11}$} \\ \hline
\multicolumn{4}{c}{System 2} \\ \hline
\multicolumn{1}{|c|}{(2,2)} &
  \multicolumn{1}{c|}{\begin{tabular}[c]{@{}c@{}}FR\\ DG\end{tabular}} &
  \multicolumn{1}{c|}{\begin{tabular}[c]{@{}l@{}}$2.1676683889035\times 10^{-6}$\\ $2.1676691548376\times 10^{-6}$\end{tabular}} &
  \multicolumn{1}{c|}{$3.5\times 10^{-7}$} \\ \hline
\multicolumn{1}{|c|}{(4,2)} &
  \multicolumn{1}{c|}{\begin{tabular}[c]{@{}c@{}}FR\\ DG\end{tabular}} &
  \multicolumn{1}{c|}{\begin{tabular}[c]{@{}c@{}}$3.7609900151243\times 10^{-11}$\\ $3.7609939802703\times 10^{-11}$\end{tabular}} &
  \multicolumn{1}{c|}{$1.1\times 10^{-6}$} \\ \hline
\end{tabular}
    \caption{Mode-by-mode energy fluxes from a circular orbit around a Schwarzschild black hole. 
    The table compares our time-domain solver results (Alg.=DG) with frequency-domain (FR) results (Alg.=FR) computed by a code made publicly available through the Black Hole Perturbation Tool Kit (BHPTK). Relative errors are computed as $|\dot{E}^{\ell m}_{\rm DG} - \dot{E}^{\ell m}_{\rm FR}| / |\dot{E}^{\ell m}_{\rm FR}|$. To showcase our code for a variety of configurations, we consider different harmonic modes and particle locations. For all modes except $(5,3)$, the particle is located at $r_{p}=10.99332834601232$, while for the $(5,3)$ we use $r_{p}=16.09469709319520$.  Results are reported for
    system 1 (our preferred first-order reduced system~\eqref{eq:wave-1p1-first-layers} that is symmetric hyperbolic) and system 2 (an alternative first-order system~\eqref{eq:teuk0-1p1-layers-first_alt} that is discussed in App.~\ref{app:alternative_system}). We find that DG schemes solving system 2 are unable to obtain accuracies that should be achievable by double precision. A complete understanding of this issue is still lacking.
    }
    \label{tab:fluxes_a0_vertical}
\end{table}

\subsubsection{Fluxes on Kerr}
\label{sec:particle-waveforms_Kerr}

We now consider the more generic Kerr case where $a\neq0$. Compared to the simpler case documented in Sec.~\ref{sec:particle-waveforms_Schwarzschild}, when $a\neq0$ infinitely many harmonic modes will mix. As discussed in Sec.~\ref{Sec:modes}, truncating the numerical computation to a maximum mode value, $\ell_{\rm max}$, will result in an additional source of error in $\dot{E}_{\ell m}$. Fortunately, the solution $\Psi$ to the Teukolsky equation is smooth at $\mathscr{I}^{+}$. And so the harmonic expansion Eq.~\eqref{eq:anstaz}, and hence $\dot{E}_{\ell m}$, will converge exponentially fast in $\ell_{\rm max}$ suggesting only the lowest $\ell$ modes will need to be retained in the computation.

\begin{figure}
    \centering
    \includegraphics[width=0.48\textwidth]{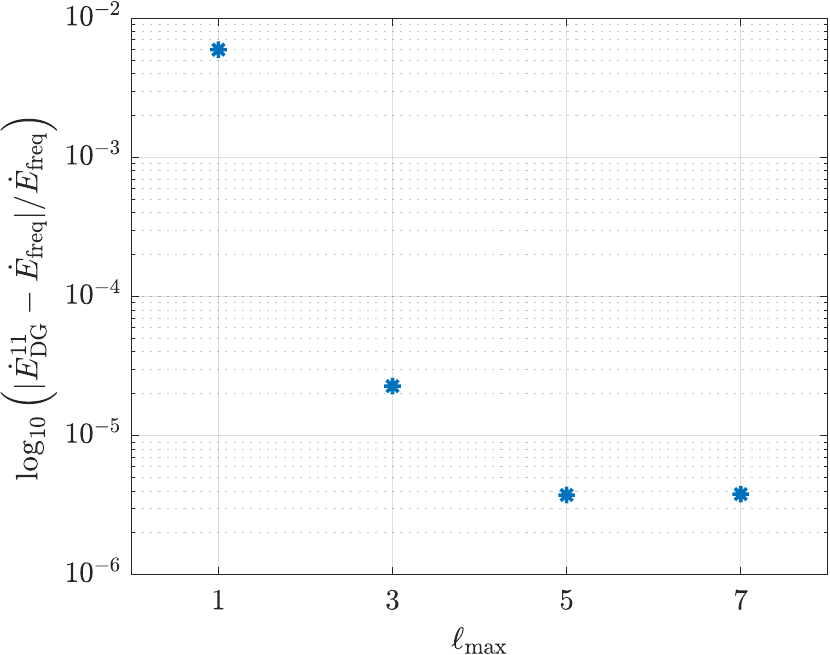}
\caption{Convergence with $\ell_{\rm max}$ in the numerically computed energy flux from a particle in a circular orbit about a Kerr black hole with spin $a=0.9$. Here, $\ell_{\rm max}$ is the largest value of $\ell$ retained in the truncated expansion~\eqref{eq:anstaz}. We have selected our grid resolution and other numerical parameters to introduce negligible error compared to the error due to mode truncation. It can be seen that the energy flux $\dot{E}^{1 1}$ converges exponentially fast with the number of harmonic modes retained. When we reach $\ell_{\rm max}=7$ we run into a numerical conditioning issue described in the text.}
\label{fig:kerr_ell_convergence}
\end{figure}

To demonstrate this, we compute $\dot{E}_{1 1}$ using both our DG method and a frequency-domain solver made available as part of the Black Hole Perturbation Toolkit~\cite{TeukFreqCode,BHPToolkit}. The frequency-domain solver includes many $\ell$ modes in its computation, and we take this value of the energy flux, $\dot{E}^{1 1}_{\rm FR}$, to be the exact one. For this experiment, we compute fluxes from a particle in a circular orbit located $r_{p}=16.0914363989845$ orbiting a Kerr blackhole with a spin of $a=0.9$. Our physical domain $\rho \in [-1600, 1600]$ is partitioned into $K=320$ subdomains and on each subdomain we approximate the solution with a degree $N=18$ polynomial. Other numerical parameters describing our setup include $\Delta t=0.017378$, $R=100$, and $s=1600$. The reason for this large grid is to keep the right boundary (where we measure the energy flux) causally disconnected from the left boundary where the correct non-reflecting boundary conditions are non-trivial. Indeed, even after enacting an azimuthal transformation~\cite{krivan1996dynamics,PhysRevD.56.3395} the $a \neq 0$ Teukolsky equation near the horizon is not the ordinary wave equation, and so simple Sommerfeld boundary conditions will lead to small, spurious reflections. Although non-reflecting boundary conditions can be derived~\cite{hagstrom2007radiation,lau2005analytic,lau2004rapid} this will be left for future work. The distributional source term is slowly turned on by setting the smoothing parameters to $\mu_{\text{smooth}}=200$ and $\delta_{\text{smooth}}=0.001$. Setting trivial initial data, we find that junk radiation has left the grid by $T=4000$M, at which time we compute $\dot{E}^{1 1}_{\rm DG}$ from our DG solver. 

Figure~\ref{fig:kerr_ell_convergence} shows exponential convergence of $\dot{E}^{1 1}_{\rm DG}$ with $\ell_{\rm max}$ and table~\ref{tab:fluxes_kerr_less_digits} provides numerical values of the energy flux as we vary the number of coupled modes and the spin parameter.
From Fig.~\ref{fig:kerr_ell_convergence} we see that convergence prematurely halts when we reach $\ell_{\rm max}=7$. We believe this is due to two compounding issues. First, as the $\psi_{71}$ is many orders of magnitude smaller than $\psi_{11}$ and is only weakly coupled to $\psi_{11}$. This suggests we need high precision to correctly capture the $(7,1)$ mode's impact on the $(1,1)$ mode. However, as described in Sec.~\ref{sec:E_at_infinity}, the expressions near $\rho=s$ appear to suffer from mild ill-conditioning for $a\neq0$ and large values of $\ell_{\rm max}$. Taken together, these two issues prevent us from seeing improved accuracy in $\dot{E}^{1 1}_{\rm DG}$ (for large values of blackhole spin) when $\ell_{\rm max}\geq 7$. We believe this issue can be fixed by using quad-precision computations, for example, although we have not considered this remedy here. Nevertheless, as far as we are aware, our Kerr energy flux computations still constitute the most accurate time-domain ones presented to date.

{\renewcommand{\arraystretch}{1.5}%
\begin{table*}[t]
\begin{tabular}{|c|c|c|c|c|}
\hline 
No. of modes coupled  & $\left(\ell,m\right)$  & Set of coupled modes & $\dot{E}_{{\rm DG}}^{\ell m}$ $\left(a=0.9\right)$ & $\dot{E}_{{\rm DG}}^{\ell m}$$\left(a=-0.9\right)$\tabularnewline
\hline 
\multirow{3}{*}{$1$} & $\left(1,1\right)$  & $\left(1,1\right)$ & $1.824775\times10^{-06}$  & $2.034678\times10^{-06}$\tabularnewline
 & $\left(3,1\right)$  & $\left(3,1\right)$ & $3.72\times10^{-12}$  & $4.62\times10^{-12}$\tabularnewline
 & $\left(5,1\right)$  & $\left(5,1\right)$ & $1.05\times10^{-18}$  & $1.46\times10^{-18}$\tabularnewline
\hline 
\multirow{2}{*}{2} & $\left(1,1\right)$  & $\left\{ \left(1,1\right),\left(3,1\right)\right\} $ & $1.835753\times10^{-06}$  & $2.047048\times10^{-06}$\tabularnewline
 & $\left(3,1\right)$  & $\left\{ \left(1,1\right),\left(3,1\right)\right\} $ & $3.74\times10^{-12}$  & $4.64\times10^{-12}$\tabularnewline
\hline 
\multirow{3}{*}{3} & $\left(1,1\right)$  & $\left\{ \left(1,1\right),\left(3,1\right),\left(5,1\right)\right\} $ & $1.835723\times10^{-06}$  & $2.047008\times10^{-06}$\tabularnewline
 & $\left(3,1\right)$  & $\left\{ \left(1,1\right),\left(3,1\right),\left(5,1\right)\right\} $ & $3.75\times10^{-12}$  & $4.66\times10^{-12}$\tabularnewline
 & $\left(5,1\right)$  & $\left\{ \left(1,1\right),\left(3,1\right),\left(5,1\right)\right\} $ & $1.06\times10^{-18}$  & $1.48\times10^{-18}$\tabularnewline
\hline 
\multirow{3}{*}{4} & $\left(1,1\right)$  & $\left\{ \left(1,1\right),\left(3,1\right),\left(5,1\right),\left(7,1\right)\right\} $ & $1.835718\times10^{-06}$  & $2.047008\times10^{-06}$\tabularnewline
 & $\left(3,1\right)$  & $\left\{ \left(1,1\right),\left(3,1\right),\left(5,1\right),\left(7,1\right)\right\} $ & $3.75\times10^{-12}$  & $4.66\times10^{-12}$\tabularnewline
 & $\left(5,1\right)$  & $\left\{ \left(1,1\right),\left(3,1\right),\left(5,1\right),\left(7,1\right)\right\} $ & $1.07\times10^{-18}$  & $1.48\times10^{-18}$\tabularnewline
\hline 
\end{tabular}
    \caption{Mode-by-mode energy fluxes from a circular orbit at radius $r_{p}=16.0914363989845$ around a Kerr black hole with spin $a=0.9$ and $a=-0.9$. We consider the energy flux $\dot{E}^{\ell m}_{\rm DG}$ as we include more coupled modes. For each $\dot{E}^{\ell m}_{\rm DG}$ computation we list the harmonic modes that have been used. We compare our results to the frequency domain code from BHPTK that, for all practical purposes, effectively includes all modes.}
    \label{tab:fluxes_kerr_less_digits}
\end{table*}

\section{Summary and future work}
\label{Sec:Summary}
This paper presents a new numerical method for solving the distributionally-forced $s=0$ Teukolsky equation (scalar waves on Kerr) that models extreme mass ratio inspirals (EMRIs) in a Kerr spacetime. Our method uses a nodal discontinuous Galerkin (DG) approach that expands the solution in spherical harmonics and recasts the sourced Teukolsky equation as a set of coupled, fully-first order one-dimensional (positive semi-definite) symmetric hyperbolic equations. This approach allows us to correctly account for the Dirac delta distribution using a modified numerical flux and achieve global spectral accuracy even at the source's location. Furthermore, we use the hyperboloidal layer method to connect the near field to future null infinity, providing direct access to the far-field waveform. Our numerical experiments demonstrate the accuracy and efficiency of the method, including convergence tests against exact solutions, energy luminosities for circular orbits, and the scheme's superconvergence at future null infinity.

Our method has several advantages over existing time-domain solvers for the distributionally-forced Teukolsky equation. First, it does not introduce small-length scales near the smaller black hole, which can be a significant source of systematic error in other time-domain solvers. Combined with spectral accuracy at the source's location, our DG scheme and code should be well-suited for self-force calculations that require highly-accuracy numerics at the location of delta distribution. Its also worth noting that when computing gravitational self-force effects a suitable regularization procedure must be used. Regularization techniques have been well-developed for the spherical harmonic modes. Yet in frequency-domain Kerr calculations, spheroidal harmonics are exclusively used, which in turn requires projecting onto a spherical harmonic basis before the regularization step can be applied. In our method, we directly solve for the spherical harmonic modes, thereby avoiding this complicated intermediate step. Second, it achieves global spectral accuracy (in fact, superconvergence), which is critical for accurately capturing the complex dynamics of extreme mass ratio inspirals both at the smaller black hole's location and in the far field. Finally, our method provides direct access to the far-field waveform, a key output of the simulation that can be compared with data from upcoming space-based gravitational wave detectors. 

As a byproduct of our work, we have compiled numerical evidence that when combined with the hyperboloidal layer method, one seemingly reasonable choice of the first-order reduction variable appears to be problematic. This issue shows up in different settings, such as calculating scalar fluxes at future null infinity or when comparing to exact solutions of the ordinary wave equation. In all settings, the numerical solution's accuracy saturates at about 5 to 6 digits irrespective of the timestep or grid resolution. On the other hand, with a better choice of auxiliary variable, we are able to achieve the expected decay of numerical error to better than 12 digits of accuracy. Appendices \ref{app:system_hyperbolicity} and \ref{app:alternative_system} discuss the hyperbolicity and well-posedness properties of both system.

There are several avenues for future work that could extend and improve our methodology. One important direction is the extension of our method to handle the $s=\pm 2$ Teukolsky equation, which are needed for modeling gravitational waves from EMRI systems. Another area for future work is the development of methods to handle eccentric orbits using a time-dependent coordinate transformation, which has been worked out in simpler settings~\cite{field2009discontinuous}. The methods presented in this paper assume circular orbits, but eccentric orbits are common in real astrophysical EMRI systems and can significantly impact the gravitational wave signal's morphology. Finally, there is also potential for further improving the accuracy and efficiency of our method by exploring higher-order time-stepping methods or adaptive mesh refinement.

\section*{Acknowledgments}
We thank Stephen Lau and Anil Zenginoglu for discussions on first-order reductions and hyperboloidal layers, Jennifer Ryan for discussions on superconvergence, and Olivier Sarbach for helpful feedback on App.~\ref{app:system_hyperbolicity}. We thank Som Bishoyi, Peter Diener, Zachariah Etienne, Alfa Heryudono, Scott A. Hughes, Harald Pfeiffer, Leo Stein, Saul Teukolsky, Alex Vano-Vinuales, Niels Warburton, and Barry Wardell for helpful discussions throughout the project. We also thank Karoly Csukas and Hannes R{\"u}ter for their valuable feedback on an earlier draft of this paper. The authors acknowledge support of NSF Grants PHY-2106755 and PHY-2307236 (G.K), DMS-1912716 and DMS-2309609 (S.F, S.G, and G.K), AFOSR Grant No. FA9550-18-1-0383 (S.G), PHY-2207780 (K.R), and Office of Naval Research/Defense University Research Instrumentation Program (ONR/DURIP) Grant No. N00014181255. MV acknowledges partial support through the cyberteams CAREER's program, NSF Grant No. OAC-2018873. K.R is a member of the Weinberg Institute and this manuscript has preprint number WI-45-2023. This material is based upon work supported by the National Science Foundation under Grant No. DMS-1439786 while a subset of the authors were in residence at the Institute for Computational and Experimental Research in Mathematics in Providence, RI, during the Advances in Computational Relativity program. Some of the simulations were performed on the UMass-URI UNITY supercomputer supported by the Massachusetts Green High Performance Computing Center (MGHPCC). We also thank ChatGPT for writing the first draft of the conclusion (and only the conclusion) of this paper.

\appendix

\section{Mode-coupling coefficients}
\label{app:coupling}

When expanding the $s=0$ Teukolsky equation~\eqref{eq:teuk0} in spherical harmonics, we encounter a
term $\sin^{2}\theta Y_{\ell m}$ that is responsible for mode coupling. To re-expand this term
with spherical harmonics
\begin{align}
\sin^{2}\theta Y_{\ell m} & = c^{\ell-2}_{\ell m}Y_{\ell-2, m} + c^{\ell}_{\ell m} Y_{\ell m} + c^{\ell+2}_{\ell m} Y_{\ell+2, m} \\
& = C_{++}^{\ell-2} Y_{\ell-2, m} + C_0^{\ell} Y_{\ell m} + C_{--}^{\ell+2} Y_{\ell+2, m} \label{eq:app1}
\end{align}
we need expressions for the mode-coupling coefficients. These can be found from either Ref.~\cite{barack2017time} or Ref.~\cite{bailey_1933}, where the latter considered coupling between associated Legendre polynomials, $P^m_{\ell}(\cos \theta)$. We follow Ref.~\cite{barack2017time}, and in line~\eqref{eq:app1} we switch from our paper's notation to the notation used by Ref.~\cite{barack2017time}. Quoting the relevant results from Ref.~\cite{barack2017time}, we have
\begin{equation} 
\begin{array}{lcl}
C_{++}^\ell &=&  -c_+^{\ell+1} c_+^\ell \,, \\
C_0^\ell    &=& 1-(c_-^\ell)^2-(c_+^\ell)^2 \,,\\
C_{--}^\ell &=&  -c_-^{\ell-1} c_-^\ell \,,
\end{array}
\end{equation}
where 
\begin{align}
c_-^\ell = \left[\frac{\left(\ell^2-m^2\right)}
                       {(2\ell-1)(2\ell+1)}\right]^{1/2} \,, \qquad
c_+^\ell = c_-^{\ell+1} \,.
\end{align}

For future use in App.~\ref{app:system_hyperbolicity}, we  summarize a few properties for the coupling coefficients 
$c^{\ell}_{\ell m} = C_0^\ell$, $c^{\ell}_{(\ell+2), m} = C_{++}^\ell$, and $c^{\ell}_{(\ell-2), m} = C_{--}^\ell$. 
The first,
\begin{align}
0 < c^{\ell}_{\ell m} < 1 \,, \qquad \lim_{\ell \rightarrow \infty, \ell=m} c^{\ell}_{\ell m} = 1 \,,
\end{align}
shows that $c^{\ell}_{\ell m}$ is positive, bounded, and approaches unity in a certain limit.
The second,
\begin{align}
c^{\ell}_{(\ell+2), m} < 0 \,, \qquad c^{\ell}_{(\ell-2), m} < 0 \,,
\end{align}
shows that the ``off-diagonal'' coupling coefficients are negative.
All of these properties follow from inspection of the relevant terms.

\section{Matrix form of the full system~\eqref{eq:teuk0-1p1-first-layers}}
\label{app:full_system}

We now consider the matrix form for the general system~\eqref{eq:teuk0-1p1-first-layers} for any combination of $\ell_{\rm max}$, $m$ and $\ell$-mode type = \{odd, even\}. The even and odd $\ell$ modes decouple so their governing systems can be analyzed separately. For concreteness, consider (for a fixed value of $m$) the even $\ell$-mode sector, where $\ell=\ell_{\rm min},\ell_{\rm min}+2,\dots,\ell_{\rm max}$. First, we split our solution vector $U_{\ell_{\rm max} m}$ into subvectors
\begin{align*}
\vec{\pi}_{\ell_{\rm max} m} =  \left[\pi_{\ell_{\rm min} m}, \pi_{(\ell_{\rm min}+2)m}, \dots, \pi_{\ell_{\rm max}m}  \right]^{T} \,, \\
\vec{\phi}_{\ell_{\rm max} m} = \left[\phi_{\ell_{\rm min}m}, \phi_{(\ell_{\rm min}+2)m}, \dots, \phi_{\ell_{\rm max}m} \right]^{T} \,,
\end{align*}
which allows us to write $U_{\ell_{\rm max} m} = \left[ \vec{\pi}_{\ell_{\rm max} m}, \vec{\phi}_{\ell_{\rm max} m} \right]$.
We now use block notation to index vectors, for example $U^{\ell_{\rm max} m}_\pi =  \vec{\pi}_{\ell_{\rm max} m}$. With this new notation, 
the matricies defining the first-order system, Eq. ~\eqref{eq:matrix_vector_system} can be partitioned as
\begin{equation}
\label{eq:E_A_matrix_as_blocks}
E = \left(
\begin{array}{c|c}
E_{\pi \pi} & E_{\pi \phi}\\
\hline
E_{\phi \pi} & E_{\phi \phi}
\end{array} \right) \,, \qquad
\hat{A} = \left(
\begin{array}{c|c}
\hat{A}_{\pi \pi} & \hat{A}_{\pi \phi}\\
\hline
\hat{A}_{\phi \pi} & \hat{A}_{\phi \phi}
\end{array} \right) \,.
\end{equation}
The block-wise components of $\hat{A}$ are $\hat{A}_{\pi \pi} = \hat{A}_{\phi \phi} = 0$ and
$\hat{A}_{\pi \phi} = \hat{A}_{\phi \pi} = \left(1-H\right) \mathbf{I} $,
where $\mathbf{I}$ is the identity matrix. 
The block-wise components of $E$ are 
$E_{\pi \phi} = E_{\phi \pi} = -H \mathbf{I}$,
$E_{\phi \phi} = \mathbf{I}$, 
and the tridiagonal matrix
\begin{widetext}
\begin{align}\label{eq:E_full}
 E_{\pi \pi}=
 \begin{pmatrix} 
 1-f c^{\ell_{\rm min}}_{\ell_{\rm min}, m} & -f c^{\ell_{\rm min}}_{\ell_{\rm min}+2, m} & 0 & 0 & 0 & 0 \\
-fc^{\ell_{\rm min}+2}_{\ell_{\rm min},m} & 1-fc^{\ell_{\rm min}+2}_{\ell_{\rm min}+2,m} & -fc^{\ell_{\rm min}+2}_{\ell_{\rm min}+4,m} & 0 & \vdots & \vdots\\
0 & -fc^{\ell_{\rm min}+4}_{\ell_{\rm min}+2,m} & 1-fc^{\ell_{\rm min}+4}_{\ell_{\rm min}+4,m} & -fc^{\ell_{\rm min}+4}_{\ell_{\rm min}+6,m}  &  0 & 0\\
0 & 0 & \ddots & \ddots & \ddots & 0\\
\vdots & \vdots & \vdots &-fc^{\ell_{\rm max}-2}_{\ell_{\rm max}-4,m} & 1-fc^{\ell_{\rm max}-2}_{\ell_{\rm max}-2,m} & -fc^{\ell_{\rm max}-2}_{\ell_{\rm max},m}\\
0 & 0 & 0 & 0  & -fc^{\ell_{\rm max}}_{\ell_{\rm max}-2,m} & 1-fc^{\ell_{\rm max}}_{\ell_{\rm max},m}
\end{pmatrix}\,.
\end{align}
\end{widetext}
The matrix structure for the odd $\ell$-modes is identical. From properties documented in App.~\ref{app:coupling}, we have $c^{\ell+2}_{\ell,m} = c^{\ell}_{(\ell+2),m}$ which implies $E_{\pi \pi}$ is symmetric. Therefore, $E$ is also symmetric. That $\hat{A}$ is symmetric follows from direct inspection. 

Finally, we consider the flux Jacobian matrix $A$ defined from Eq.~\eqref{eq:matrix_vector_system2}. In terms of block-wise components we have
\begin{equation} \label{eq:A_matrix} 
\frac{A}{1-H} = 
\left(
\begin{array}{c|c}
H \left(E_{\pi \pi} - H^2 \mathbf{I}\right)^{-1} & \left( E_{\pi \pi} - H^2\mathbf{I} \right)^{-1}  \\
\hline
\left( \mathbf{I} - H^2 E_{\pi \pi}^{-1} \right)^{-1} & H\left( \mathbf{I} - H^2 E_{\pi \pi}^{-1} \right)^{-1} E_{\pi \pi}^{-1}
\end{array} \right) \,,
\end{equation}
where $E_{\pi \pi}^{-1} E_{\pi \pi} = \mathbf{I}$. We see that, in general, $A$ is not a symmetric matrix, even though $E_{\pi \pi}^{-1}$ is. By considering $H=0$ we see that each of the four blocks of $A$ are symmetric when working in the original $t$ and $r_{*}$ coordinates. As a special case, $A$ is symmetric when $a=0$.

\section{The first-order Eq.~\eqref{eq:teuk0-1p1-first-layers} is a positive (semi-definite) symmetric hyperbolic system}
\label{app:system_hyperbolicity}

With an explicit expression for the matricies of our problem given in App.~\ref{app:full_system}, we now show conditions under which the first-order system~\eqref{eq:teuk0-1p1-first-layers} is symmetric hyperbolic. Recall that to study a first-order system's hyperbolicity classification it is sufficient to consider the principle part of the differential equations,
\begin{align}
E\dot{U}+\hat{A}U^{\prime}+ \dots &= 0 \,, \label{eq:pp_type1} \\
\dot{U}+AU^{\prime}+ \dots &=0 \,, \label{eq:pp_type2}
\end{align}
where the variable-coefficient matrices $A(\rho)$, $E(\rho)$, and $\hat{A(\rho)}$ depend on the spatial independent variable $\rho$ and ``$\dots$'' denotes lower-order terms (ie. it does not include any term proportional to $\dot{U}$ or $U^{\prime}$). Eq.~\eqref{eq:pp_type1} represents the form of our original system~\eqref{eq:teuk0-1p1-first-layers}'s principle part after we put it into matrix form~\eqref{eq:matrix_vector_system} where the matrices are given in App.~\ref{app:full_system}. System~\eqref{eq:pp_type2} is the result of inverting $E$, leading to the system~\eqref{eq:matrix_vector_system2} that we discretize with the DG method.

One important question is whether or not this system constitutes a well-posed initial-boundary-value problem (IBVP). That is, given sufficiently smooth initial conditions $U(0,\rho)$ and dissipative boundary conditions, the $L_2$ norm of $U$ can be bounded by norms taken over the initial data and boundary conditions~\cite{sarbach2012continuum,majda1975initial,gustafsson1995time}. Well-posedness for linear, variable-coefficient systems like ours can be shown if the problem is symmetric hyperbolic system~\cite{gustafsson1995time}. This condition is sometimes stated as follows: the system~\eqref{eq:pp_type2} is symmetric hyperbolic if $A(\rho)$ is a symmetric matrix at all values of $\rho$ in the physical domain~\cite{gustafsson1995time}. Direct inspection of Eq.~\eqref{eq:A_matrix} shows that this is not the case. However, a similar well-posedness result follows if we show our system~\eqref{eq:pp_type1} to be positive symmetric: that is both $E(\rho)$ and $\hat{A}(\rho)$ are symmetric matrices at all values of $\rho$, and $E(\rho)$ is a positive-definite matrix~\footnote{Recall that $E$ is positive definite if $z^T E z > 0$ for all non-trivial vectors $z$. A matrix $E$ is positive semi-definite if $z^T E z \geq 0$ for all non-trivial vectors $z$} at all values of $\rho$. To see why this would be helpful, note that if $E$ is positive definite then we can take its square root, $E = C C$, for some invertible matrix $C$. Defining a new system vector $V = C U$, Eq.~\eqref{eq:pp_type1} can be rewritten as
\begin{align} \label{eq:pp_type3}
\dot{V}+C^{-1}\hat{A}C^{-1}V^{\prime}+ \dots &= 0 \,,
\end{align}
where the matrix $C^{-1}\hat{A}C^{-1}$ is symmetric~\cite{courant1962methods}. System~\eqref{eq:pp_type3} is clearly symmetric hyperbolic by the usual definition, and so the standard well-posedness results hold for $V$. Due the boundedness of $C$ and $C^{-1}$, the original system~\eqref{eq:pp_type1} is also well-posed. 

As we show below, $E(\rho)$ is positive definite everywhere in the physical domain except at the right-most boundary point
$\rho=s$ where $E(\rho)$ becomes positive semi-definite. And so $C$ is singular at this location. To proceed, let us define a system energy 
\begin{align}
\label{eq:energy_E}
{\cal E}(\tau) & = \frac{1}{2} \int U^T E U d\rho  = \frac{1}{2} \int V^T V d\rho \,,
\end{align}
where $U^T$ is the transpose of $U$. Then ${\cal E}(\tau) \geq 0 $ due to $E(\rho)$ being positive (semi-)definite.
Our approach is as follows: provided the equality between norms of $U$ and $V$ in Eq.~\eqref{eq:energy_E} hold, the problem for $U$ is well-posed if we show the problem for $V$ is well-posed. Well-posedness for $V$
follows from it being symmetric hyperbolic: we show the system for $U$ is positive symmetric at all points in the physical domain except at the right-most boundary point $\rho=s$ (from which it follows the system for $V$ is symmetric hyperbolic except at $\rho=s$), and at $\rho=s$ the system for $V$ is symmetric hyperbolic by direct inspection. 
Caveats about strong well-posedness, which are applicable to the first-order system~\eqref{eq:teuk0-1p1-first-layers} as well, are summarized in App.~\ref{app:singular_boundary_matrix}.

\subsection{Setup}

The system of first-order differential equations~\eqref{eq:teuk0-1p1-first-layers} are solved on an open spatial domain, where $0 \leq a \leq 1$ and $a=1$ corresponds to an extremal Kerr blackhole. In the original $(t,r_*)$ coordinate system ($\Omega=1$ and $H=0$) this is $r_* \in [\rho_L, \infty)$ and we note that $r_*=-\infty$ is the black hole's horizon. In hyperboloidal coordinates $(\tau,\rho)$ the domain is $\rho \in [\rho_L, s]$ where $r_*(\rho=s) = \infty$. 

Let $D = [\rho_L, s]$ be the full domain of interest and $D\setminus\{s\}$ the domain after omitting the point at infinity. The reason for partitioning the domain in this way is because on $D\setminus\{s\}$ the matrix $E$ is positive definite while at $\rho=s$ its positive semi-definite. We will return to $\rho=s$ in App.~\ref{app:System_at_infinity}.

Since App.~\ref{app:full_system} shows that $E$ and $\hat{A}$ are symmetric, we only need to show that $E$ is positive definite. While we have been unable to prove $E$ is positive definite in the most general setting, we have been able to prove if for some important cases: (i) $a=0$ and any allowable function $H$ for the problem defined on $\rho \in D\setminus\{s\}$, (ii) $a\neq0$ and $H=0$ for the problem defined on $\rho \in D$, and (iii) with $a\neq0$ and any allowable function $H$ for the problem defined on $\rho \in D\setminus\{s\}$ one simply needs to check the whether the inequality~\eqref{eq:diag_dom_condition_E} is satisfied. This inequality can be checked for any specific $H$ and could be used to help select suitable ones. 

\subsection{$a=0$ with hyperboloidal layers}

We first consider a special case: when $a=0$ (scalar waves on Schwarzschild) the matrix $E$ is positive definite for all $\rho \in D\setminus\{s\}$. Upon further setting $M=0$, the results of this subsection also apply to the ordinary flatspace wave equation in hyperbolodial coordinates~\eqref{eq:wave-1p1-first-layers}.

When $a=0$ this implies $f=0$, and the matrix $E_{\pi \pi} = \mathbf{I}$ vastly simplifies. In this case we can directly check the positive-definite condition. Mimicking the block structure of $E$, let 
$z = \left[a_1, a_2, \dots, a_n, b_1, b_2, \dots, b_n \right]$, such that $a_i$ and $b_i$ correspond, respectively, to the $\pi$ and $\phi$ portions of $U$. Direct computation gives 
\begin{align*}
z^T E z = \sum_{i=1}^n a_i^2 + b_i^2 - 2 H a_i b_i \,.
\end{align*}
Noting that $0 \leq H(\rho) < 1$ for $\rho \in D\setminus\{s\}$, we have two cases to consider for each $i^\mathrm{th}$ term. First suppose $a_i b_i <0$, then clearly $a_i^2 + b_i^2 - 2 Ha_i b_i \geq 0$. Next consider $a_i b_i >0$, in which case we have
\begin{align*}
a_i^2 + b_i^2 - 2 Ha_i b_i >  a_i^2 + b_i^2 - 2 a_i b_i \geq 0 \,
\end{align*}
where we've used Young's inequality, $ab \leq a^2/2 + b^2/2$. Such bounds can be applied to all terms in the sum, and so we have $z^T E z > 0$. Finally, we note that at $\rho=s$ we have $H=1$, and repeating the steps above we find $E$ to be positive semi-definite as $z^T E z \geq 0$. 

\subsection{The general case}

We now consider a more general setting by allowing $a\neq0$, and our proof follows a different strategy than the previously considered $a=0$ case. In particular, because  $E$ is symmetric and real, then $E$ is positive definite if (i) the diagonal entries of $E$ are positive and (ii) $E$ is diagonally dominant. We derive the conditions under which matrix $E$ is positive definite for all $\rho \in D\setminus\{s\}$.

We first show the diagonal entries are positive, which in turn requires that the diagonal entries of both $E_{\phi \phi}$ and $E_{\pi \pi}$ are positive. Because $E_{\phi \phi} = \mathbf{I}$ this is obviously true. Next we consider the diagonal of $E_{\pi \pi}$, whose elements are of the form $1-f c^{\ell}_{\ell m}$ where $f(r;a)$ is defined in Eq.~\eqref{eq:def_mu_f}. Let us first remark that $\Delta >0$ (and hence $f>0$) outside of the horizon, which our computational grid is restricted to. 
From App.~\ref{app:coupling} recall that $0 < c^{\ell}_{\ell m} < 1$, and so $1-f < 1-f c^{\ell}_{\ell m}$. At this point, we use Mathematica to find an upper bound on $f(r;a)$. We do this by first finding $r=r_\mathrm{max}$ that maximizes $f(r;a)$ for a fixed value of $a$. Setting $\partial_r f = 0$, we find 
$r_\mathrm{max}$ solves the cubic equation $r^3 - 3Mr^2 + a^2r + a^2M=0$, which has one real root outside of the horizon. We now view $f(r=r_\mathrm{max};a)$ as a function of $a$. Numerical evidence (and physical intuition) suggests that this function is maximized at $a=1$. In fact, for $a=M=1$ we have $r_\mathrm{max}=1+\sqrt{2}$. Evaluating we get 
\begin{align}
\max_{\rho \in D, 0\leq a \leq 1} f(r;a) = f(1+\sqrt{2};1)=3/4 - 1/\sqrt{2} \approx 0.043 \,.
\end{align}
Therefore 
\begin{align}
0 < 1/4 + 1/\sqrt{2} \leq 1-f < 1-f c^{\ell}_{\ell m}\,,
\end{align}
over all allowable values of $a$ and $r$. We conclude that the diagonal entries $1-f c^{\ell}_{\ell m}$ are all positive.

Next we show $E$ is strict diagonally dominant. Recall that a matrix is said to be strict diagonally dominant if 
\begin{align*}
\sum_{j \neq i} |e_{ij}| < |e_{ii}|  \qquad \forall i \,,
\end{align*}
where $e_{ij}$ is the entry located at the $i^{th}$ row and $j^{th}$ column of the matrix $E$. For the matrix $E$ to be
strict diagonally dominant, the size of its $i^{th}$ diagonal element must be larger than the sum of all the non-diagonal entries in the $i^{th}$ row.

Considering the block structure of our matrix, we first inspect the rows corresponding to $\left( E_{\phi \pi} | E_{\phi \phi} \right)$. The diagonals are all $e_{ii} = 1$ while the sum of the entries in row $i$ is $\sum_{j \neq i} |e_{ij}| = |H| = H$. Noting that $0 \leq H < 1$ for $\rho \in D\setminus\{s\}$, the strict diagonally-dominant condition is satisfied for the bottom half rows of $E$. At $\rho=s$, $H=1$ and $E$ satisfies the weak diagonally-dominant condition $\sum_{j \neq i} |e_{ij}| \leq |e_{ii}|$.

Finally, we consider the rows corresponding to
$\left( E_{\pi \pi} | E_{\pi \phi} \right)$. Using previously stated properties of the coupling coefficients, $f$, and $H$, the condition we need to check for strict diagonal dominance is 
\begin{align}\label{eq:diag_dom_condition_E}
c^{\ell}_{\ell m} - c^{\ell}_{(\ell+2), m} - c^{\ell}_{(\ell-2), m} \stackrel{?}{<} \frac{1-H}{f} \,.
\end{align}
The left-hand-side can be evaluated for any value of $(\ell,m)$.
A brute force (numerical)
search up to $m=\ell=3000$ reveals that
$c^{\ell}_{\ell m} - c^{\ell}_{(\ell+2), m} - c^{\ell}_{(\ell-2), m}$ is 
maximized at $(\ell,m)=(2,0)$ with a value of $1.029883$. The right-hand-side depends on $H$, which in turn 
depends on the functional form of the compression factor $\Omega$. And so we first consider $H=0$, which corresponds to 
the original $(t,r_{*})$ coordinate system. Note that the maximum of $f(r,a)$ (overall allowable values or $r$ and $a$)
was previously found to be $\approx 0.043$, we find the inequality~\eqref{eq:diag_dom_condition_E} is satisfied. Hence $E$
is positive definite for all $0\leq a \leq 1$ whenever $H=0$. The system in original $(t,r_*)$ coordinates (where $H=0$) is positive symmetric hyperbolic everywhere. 

Now consider the hyperboloidal layer transformation discussed in
Sec.~\ref{Sec:Layers}. The obvious complication is the large variety of possible functions $H$ one may encounter: both the functional form of the compression function $\Omega$ as well as any relevant parameters. For example, for the function used in this paper~\eqref{eq:compression_and_height}, the compression function has parameters $s$, $R$, and $P$.
We can bring together the previous discussion to provide an inequality that one may check once a compression function $\Omega$ is selected:
\begin{align}
1.029883 \stackrel{?}{<} \frac{1-H}{f} = \frac{\Omega^{2}}{(\Omega - \rho \Omega^{\prime})f} \,.
\end{align}
That is, $E$ is strict diagonally dominant for any $0\leq a \leq 1$ provided this inequality holds throughout the domain $\rho \in D$. For the cases we've checked (e.g. using $P=4$ and values of $R$ and $s$ similar to those used in our numerical experiments section), this inequality is always satisfied and the value of $\frac{1-H}{f}$ is typically about 10.

\subsection{System at infinity}
\label{app:System_at_infinity}

As shown above, when using hyperboloidal coordinates the matrix $E$ is positive semi-definite at $\rho=s$. In practice this means that while we still have $E = C C$ and can form $V = C U$, the matrix $C$ is not invertible. Nevertheless, by considering the structure of the relevant matricies at $\rho=s$ we will show its still possible to put the system into the form of Eq.~\eqref{eq:pp_type3}. In particular, this means the system~\eqref{eq:pp_type3} is well-defined and symmetric hyperbolic everywhere in the physical domain $D$. We also show that the matrix $A = E^{-1} \hat{A}$ is symmetric at $\rho=s$. 

At $\rho=s$ we have $H=1$, although for now we will retain $H$ in our expressions. The relevant matrices in block form~\eqref{eq:E_A_matrix_as_blocks} evaluated at $\rho=s$ are
\begin{equation}
\label{eq:E_A_matrix_as_blocks_infinity}
E = \left(
\begin{array}{c|c}
\mathbf{I} & -H \mathbf{I}\\
\hline
-H \mathbf{I} & \mathbf{I}
\end{array} \right) \,, \qquad
\hat{A} = \left(
\begin{array}{c|c}
0 & (1-H)\mathbf{I}\\
\hline
(1-H)\mathbf{I} & 0
\end{array} \right) \,,
\end{equation}
while from Eq.~\eqref{eq:A_matrix} we have
\begin{equation} \label{eq:A_matrix_inf} 
A = \frac{1}{1+H}
\left(
\begin{array}{c|c}
H\mathbf{I} & \mathbf{I}  \\
\hline
\mathbf{I} & H\mathbf{I}
\end{array} \right) \,.
\end{equation}
The matrix $A$ is clearly well-defined and symmetric, despite that fact that $E$
is not invertible at $\rho=s$.

Finally we show that $C^{-1}\hat{A}C^{-1}$ is well-defined despite the 
fact that $C$ is is not invertible at $\rho=s$. To show this, we will show the 2-norm
$\| C^{-1}\hat{A}C^{-1} \|_2$ is bounded. We first recall that given some matrix $M$, 
the 2-norm of $M$
is equal to the maximum the square root of the largest eigenvalue of $M^T M$, that is
\begin{align*}
\| M \|_2^2 = \max_i \lambda_i(M^T M) \,,
\end{align*}
where $\lambda_i(M^T M)$ is the $i^\mathrm{th}$ eigenvalue of $M^T M$. We now compute
\begin{align*}
\| C^{-1}\hat{A}C^{-1} \|_2^2 & = \max_i \lambda_i( (C^{-1}\hat{A}C^{-1})^T C^{-1}\hat{A}C^{-1}) \\
& = \max_i \lambda_i( C^{-1}\hat{A}C^{-1} C^{-1}\hat{A}C^{-1}) \\
& = \max_i \lambda_i( C^{-1}\hat{A} E^{-1} \hat{A}C^{-1}) \\
& = \max_i \lambda_i( C^{-1}\hat{A} A C^{-1}) \\
& = \max_i \lambda_i( C A A C^{-1}) \\
& = \max_i \lambda_i( A A ) = \| A \|_2^2 \,.
\end{align*}
And so the norms of $C^{-1}\hat{A}C^{-1}$ and $A$ are identical
at every point $\rho \in D$. In particular, 
at $\rho=s$ we have that $C^{-1}\hat{A}C^{-1}$ is bounded
because $A$ is bounded. Therefore $C^{-1}\hat{A}C^{-1}$ exists, and 
hence the associated system~\eqref{eq:pp_type3} is symmetric hyperbolic 
everywhere in the the physical domain of interest, including $\rho=s$.

As shown in Fig.~\ref{fig:convergence_at_scri}, our preferred system (which arises from using hyperboloidal layers with the reduction variables $\pi_{\ell m}=-\partial_{t} \psi_{\ell m}$ and $\phi_{\ell m}=\partial_{r*} \psi_{\ell m}$) exhibits the spectral convergence to round-off. Meanwhile, if we had instead used $\pi_{\ell m}=-\partial_{\tau} \psi_{\ell m}$ and $\phi_{\ell m}=\partial_{\rho} \psi_{\ell m}$ as our first-order reduction variables, does not show the expected spectral convergence. The next appendix considers this system in more detail.

\section{An alternative first-order reduction}
\label{app:alternative_system} 

Throughout the main body of this paper, we have worked with a first-order
reduction of Eq.~\eqref{eq:teuk0-1p1-v3} that arise from the reduction variables
$\pi_{\ell m}=-\frac{\partial\psi_{\ell m}}{\partial t}$
and $\phi_{\ell m}=\frac{\partial\psi_{\ell m}}{\partial r_*}$ followed by 
a coordinate transformation~\ref{eq:hyperboloidal_transformation}. 
Appendix~\ref{app:system_hyperbolicity} shows the conditions under which the resulting system 
is positive (semi-definite) symmetric hyperbolic.

Different choices in our first-order
reduction can lead to different systems with different hyperbolicity classifications. In this 
appendix we consider a reasonable set of choices that leads to a strongly hyperbolic system
but not symmetric hyperbolic.
Through numerical experiments, and as documented in Sec.~\ref{Sec:Experiments}, 
we found this formulation to be problematic leading to a loss of convergence. 
While we lack a full understanding of the issue, we note that symmetric hyperbolic systems 
are automatically well-posed IBVPs.
On the other hand, strongly hyperbolic systems need not be well-posed when one of the
wave speeds is $0$ on the boundary~\cite{sarbach2012continuum,majda1975initial}.
We note that at the rightmost boundary, $\rho=s$, one of the system's wavespeeds is $0$.
However, we caution the reader that because we do not understand the observed problematic 
behavior, we cannot entirely rule out more pedestrian explanations like a code bug. However,
we did spend significant effort checking the correctness of our code and carefully considered
issues such as catastrophic cancellation that might lead to loss of accuracy near $\rho=s$. 

Using the hyperboloidal layer approach described in Sec.~\ref{Sec:Layers}, 
the 1+1 scalar wave Eq.~\eqref{eq:teuk0-1p1-v3} becomes
\begin{align}
\label{eq:teuk0-1p1-v3-layers_alt}
& -\left(1-H^2 \right) \ddot{\psi}_{\ell m} + \left(1-H\right)^2 \psi^{\prime \prime}_{\ell m} 
+ \frac{\Delta a^2}{(r^2+a^2)^2}  c^{\ell}_{L m} \ddot{\psi}_{L m} \nonumber \\
& - 2H\left(1-H\right) \dot{\psi}^{\prime}_{\ell m} - H^{\prime}\left(1-H\right) \dot{\psi}_{\ell m} - H^{\prime}\left(1-H\right) \psi_{\ell m}^{\prime} \nonumber \\
& - \frac{4\mathrm{i}mMar}{(r^2+a^2)^2}\dot{\psi}_{\ell m} + V_{\ell m} (r)\psi_{\ell m} = g_{\ell m}(t,r) \,,
\end{align}
where we use an over-dot to denote $\partial / \partial_{\tau}$ differentiation, 
a prime for differentiation by $\partial / \partial_{\rho}$, and
the various other quantities appearing in this equation have been defined in Sec.~\ref{Sec:Layers}.
To the left of the layer, where $\rho < R$, we have $H=0$ and so both Eq.~\eqref{eq:teuk0-1p1-v3}
and Eq.~\eqref{eq:teuk0-1p1-v3-layers_alt} are identical. 
The coefficients of each term in Eq.~\eqref{eq:teuk0-1p1-v3-layers_alt}
are finite at $\rho=s$, which can be seen by noting that as $\rho \rightarrow s$ we have
$\left(1 - H\right) \sim \Omega^2 \sim r_*^{-2} \sim r^{-2}$ ~\cite{zenginouglu2011hyperboloidal,zenginouglu2011null}.

To carry out a first-order reduction, we introduce two new variables, 
\begin{align} \label{eq:bad_reduction_vars}
 & \pi_{\ell m}=-\frac{\partial\psi_{\ell m}}{\partial\tau} \,, \qquad \phi_{\ell m}=\frac{\partial\psi_{\ell m}}{\partial\rho}\,.
\end{align}
The following first-order system corresponds to the original second-order wave equation~\eqref{eq:teuk0-1p1-v3-layers_alt}:
\begin{widetext}
\begin{subequations}\label{eq:teuk0-1p1-layers-first_alt}
\begin{align}
& \dot{\psi}_{\ell m} = -\pi_{\ell m} \\
& \left(1-H^2\right) \dot{\pi}_{\ell m} 
- \frac{\Delta a^2}{(r^2+a^2)^2} c^{\ell}_{Lm} \dot{\pi}_{L m} 
- H \left(1-H\right) \dot{\phi}_{\ell m} 
 = - \left(1-H\right)^2 \phi^{\prime}_{\ell m} 
- H \left(1-H\right)  \pi^{\prime}_{\ell m}  \nonumber \\
& - H^{\prime} \left(1-H\right) \pi_{\ell m}
- \frac{4\mathrm{i}mMar}{(r^2+a^2)^2}\pi_{\ell m}
+ H^{\prime} \left(1-H\right) \phi_{\ell m}
- V_{\ell m} (r)\psi_{\ell m} + g_{\ell m}(t,r)
 \\
& \dot{\phi}_{\ell m}  = - \partial_\rho  \pi_{\ell m} \,.
\end{align}
\end{subequations}
\end{widetext}
where in carrying out the first-order reduction, we replaced the mixed-partial derivative term 
as $2 \dot{\psi}^{\prime}_{\ell m} = -\partial_\rho \pi_{\ell m} + \partial_\tau \phi_{\ell m}$ and did not consider 
alternative choices. For example, the choice $2 \dot{\psi}^{\prime}_{\ell m} = -2\partial_\rho \pi_{\ell m}$ will lead to a simpler
$A$ matrix (cf.~\ref{eq:matrix_vector_system}) but was not considered in our first-order reduction. If 
$\psi_{\ell m}$ solves the first-order system~\eqref{eq:teuk0-1p1-layers-first_alt} it also solves the 
original second-order equation~\eqref{eq:teuk0-1p1-v3-layers_alt} provided $ \phi_{\ell m}=\partial_\rho \psi_{\ell m}$.

Let us now put Eq.~\eqref{eq:teuk0-1p1-layers-first_alt} into matrix form~\eqref{eq:matrix_vector_system}. 
For concreteness, consider the even $\ell$-mode sector with $m=0$ and $\ell_{\rm max}=2$. 
Then we have $U=\left[\pi_{00}, \pi_{20}, \phi_{00}, \phi_{20}\right]^{T} $, and
\begin{align}
E=\begin{bmatrix}
1-H^{2}-fc_{00}^0 & -fc_{20}^0 & -\omega H & 0\\
-fc_{00}^2 & 1-H^{2}-fc_{20}^2 & 0 & -\omega H\\
0 & 0 & 1 & 0\\
0 & 0 & 0 & 1
\end{bmatrix}\end{align}
and
\begin{align}
\hat{A}=\begin{bmatrix}\omega H & 0 & \omega^{2} & 0\\
0 & \omega H & 0 & \omega^{2}\\
1 & 0 & 0 & 0\\
0 & 1 & 0 & 0
\end{bmatrix}
\end{align}
with 
\begin{align}
\hat{B}=\begin{bmatrix}\omega H^{\prime}+\mu & 0 & \omega\omega^{\prime} & 0\\
0 & \omega H^{\prime}+\mu & 0 & \omega\omega^{\prime}\\
0 & 0 & 0 & 0\\
0 & 0 & 0 & 0
\end{bmatrix}
\end{align}
and,
\begin{equation}
\hat{V}=\begin{bmatrix}V_{0}\psi_{0}, & V_{2}\psi_{2}, & 0, & 0\end{bmatrix}^{T}.
\end{equation}
where $\mu$ and $f$ have been defined in Eq.~\eqref{eq:def_mu_f}.

We immediately see that 
$\hat{A}$ is not symmetric, nor is $E^{-1} \hat{A}$ symmetric.
However, one can show strong hyperbolicity and, noting that at least one of the wavespeeds is zero at $\rho=s$ (the right-most boundary),
we see that $\hat{A}$ is not invertible.

\subsection{Potential issues with a singular boundary matrix}
\label{app:singular_boundary_matrix}

Hyperbolic initial-boundary problems (IBVPs) are typically called well-posed if a (weighted) $L_2$ norm of $U$, for example the energy ${\cal E}(\tau)$ define in Eq.~\eqref{eq:energy_E}, can be appropriately bounded by the initial data and boundary conditions~\cite{sarbach2012continuum,majda1975initial,gustafsson1995time}. The IBVP is called {\em strongly} well-posed if one can also control the norm of the solution (and its regularity) in a neighborhood of the boundary; see Eq.~(179) of Ref.~\cite{sarbach2012continuum}.

Following the notation of Ref.~\cite{sarbach2012continuum}, and noting that our system has one spatial dimension, the right boundary matrix is given by $P = A(\rho = s)$. Notice that some of the eigenvalues of the boundary matrix are zero. Hence, the boundary matrix is singular and, as a result, we {\em cannot} immediately conclude strong well-posedness even for symmetric hyperbolic systems~\cite{sarbach2012continuum}.

Numerically, we have been unable to achieve good results with the system arising from reduction variables~\eqref{eq:bad_reduction_vars}. For example, in the very simple case of $M=0$ (ie flatspace) the numerical solution to this system stops converging after about 4 to 5 digits of accuracy. This is shown most clearly in Fig.~\ref{fig:convergence_at_scri}: compare ``System 2'' (the problematic one using reduction variables~\eqref{eq:bad_reduction_vars}) and ``System 1'' (the first-order system that arises from the reduction variables~\eqref{eq:first_order_aux_vars_good}). A singular boundary matrix could explain these empirical observations and further analysis following Refs.~\cite{ohkubo1981regularity,rauch1985symmetric,secchi1995linear} could be considered in future work. However, as mentioned above and repeated here, we caution the reader that because we do not understand the observed problematic behavior, we cannot rule out more pedestrian explanations like a code bug despite carefully checking our code.

\bibliography{References}

\end{document}